\documentclass[12pt]{iopart}

\usepackage{epsfig}
\usepackage[dvips]{color}

\def\alt{\raise0.3ex\hbox{$\;<$\kern-0.75em\raise-1.1ex\hbox{$\sim\;$}}}
\def\agt{\raise0.3ex\hbox{$\;>$\kern-0.75em\raise-1.1ex\hbox{$\sim\;$}}}
\def\d{{\rm d}}
\newcommand{\be}{\begin{equation}}
\newcommand{\ee}{\end{equation}}
\newcommand{\bea}{\begin{eqnarray}}
\newcommand{\eea}{\end{eqnarray}}
\newcommand{\vv}{\,\,\, ,}
\newcommand{\pp}{\,\,\, .}

\definecolor{Black}{named}{Black}
\definecolor{Red}{named}{Red}

\begin{document}

\hfill DSF-24/2006, IFIC/06-21, MPP-2006-89

\title[Ultra High Energy Neutrinos in the Mediterranean]{Ultra High Energy
Neutrinos in the Mediterranean: detecting $\nu_\tau$ and $\nu_\mu$
with a km$^3$ Telescope}

\author{A. Cuoco$^1$, G. Mangano$^1$, G. Miele$^{1}$, S. Pastor$^2$,\\
L. Perrone$^3$, O. Pisanti$^1$, and P.D. Serpico$^4$}

\address{$^1$ Dipartimento di Scienze Fisiche, Universit\`{a} di Napoli
$Federico$ $II$ and INFN Sezione di Napoli, Complesso
Universitario di Monte S.\ Angelo, Via Cinthia, I-80126 Napoli,
Italy.}

\address{$^2$ Instituto de F\'{\i}sica Corpuscular (CSIC-Universitat de
Val\`{e}ncia), Ed.\ Institutos de Investigaci\'{o}n, Apdo.\ 22085,
E-46071 Val\`{e}ncia, Spain.}

\address{$^3$ Dipartimento di Ingegneria
dell'Innovazione, Universit\`{a} di Lecce and INFN Sezione di
Lecce, Via per Monteroni, I-73100 Lecce, Italy.}

\address{$^4$ Max-Planck-Institut f\"{u}r Physik
(Werner-Heisenberg-Institut), F\"{o}hringer Ring 6, D-80805
Munich, Germany.}

%%%%%%%%%%%%%%%%%%%%%%%%%%%%%%%%%%%%%%%%%%%%%%%%%%%%%%%%%%%%%%%%%%%%%%%%
%%%%%%%%%%%%%%%........... Abstract .................%%%%%%%%%%%%%%%%%%%
%%%%%%%%%%%%%%%%%%%%%%%%%%%%%%%%%%%%%%%%%%%%%%%%%%%%%%%%%%%%%%%%%%%%%%%%
\begin{abstract}
We perform a study of the ultra high energy neutrino detection
performances of a km$^3$ Neutrino Telescope sitting at the three
proposed sites for \verb"ANTARES", \verb"NEMO" and \verb"NESTOR"
in the Mediterranean sea. We focus on the effect of the underwater
surface profile on the total amount of yearly expected $\tau$ and
$\mu$ crossing the fiducial volume in the limit of full detection
efficiency and energy resolution. We also emphasize the possible
enhancement of matter effect by a suitable choice of the geometry
of the Telescope.
\end{abstract}

%    Neutrino, muon, pion, and other elementary particles; cosmic rays
%    Neutrino, muon, pion, and other elementary particle detectors; cosmic ray detectors
%    Neutrino interactions

\maketitle

\section{Introduction}

Neutrinos are one of the main components of the cosmic radiation in
the ultra-high energy (UHE) regime.  Although their fluxes are
uncertain and depend on the production mechanism, their detection
can provide information on the sources and origin of the UHE cosmic
rays. For example, UHE neutrinos can be produced via
$\pi$-photo\-production by strongly accelerated hadrons in presence
of a background electromagnetic field.  This scenario is expected to
occur in extreme astrophysical environments like the jets of active
galactic nuclei, radio galaxies and gamma ray burst sources as well
as in the propagation of UHE nucleons scattering off the cosmic
background radiation (known as {\it cosmogenic neutrinos}
\cite{[1],[2]}).

{}From the experimental point of view, after the first pioneering
and successful achievements, neutrino astronomy in the high energy
regime \cite{[3],[4],[5],[6],[7]} is a rapidly developing field,
with a new generation of neutrino telescopes on the way. A
benchmark result was obtained by the \verb"DUMAND" \cite{[8]}
collaboration, followed by the successful deployments of
\verb"NT-200" at Lake Baikal \cite{[9]} and \verb"AMANDA"
\cite{[10]} at the South Pole, which have shown the feasibility of
large optical Cherenkov neutrino telescopes (NT) in open media
like sea- or lake-water and glacial ice. These experiments
observed atmospheric neutrinos \cite{[11]} and set bounds on their
extraterrestrial flux \cite{[12],[13],[14]} which are much more
constraining than the corresponding bounds obtained by underground
neutrino detectors \cite{[15]}. These interesting results and the
perspective to perform astronomical studies using UHE neutrinos
stimulated several proposals and R\&D projects for neutrino
telescopes in the deep water of the Mediterranean sea, namely
\verb"ANTARES" \cite{[16]}, \verb"NESTOR"
\cite{Anassontzis:1999gp,Aggouras:2004mh,Aggouras:2005bg} and
\verb"NEMO" \cite{[18]}, which in the future could lead to the
construction of a km$^3$ telescope as pursued by the \verb"KM3NeT"
project \cite{[19],[20]}. Actually, the \verb"ANTARES"
collaboration is in a more advanced phase, with a telescope with
an area of $\sim$\,0.1 km$^2$  already under construction
\cite{Aguilar:2006rm}. A further project is \verb"IceCube", a
cubic-kilometer under-ice neutrino detector \cite{[21],[22],[23]}
currently being deployed in a location near the geographic South
Pole in Antarctica. \verb"IceCube" applies and improves the
successful technique of \verb"AMANDA" to a larger volume.

Until now, the possibility to perform astronomy with neutrinos has
been seriously limited by the presence of the heavy atmospheric
background in the energy range presently explored. The start of
$\nu$-astronomy is eagerly waiting the completion of the new
km$^3$ projects. In fact, according to theoretical expectations,
the km$^3$ is the minimum detector size required to detect with
reasonable chances of success point-like sources at the TeV scale
and, more relevant for this paper, to explore energies above about
100 TeV, where extraterrestrial diffuse fluxes should start to
dominate over the steeper atmospheric spectrum. In the following
we shall focus the attention mainly on this energy range, although
many of our results are valid also at lower energies.

Although NT's were originally thought as $\nu_{\mu}$ detectors,
their capability as $\nu_{\tau}$ detectors has become a hot topic
\cite{Gandhi:1995tf,[24],[25],Anchordoqui:2005is,Yoshida:2003js,Beacom:2003nh,Athar:2000rx,Bugaev:2003sw,Ishihara},
in view of the fact that neutrino oscillations lead to nearly equal
astrophysical fluxes for the three neutrino flavors\footnote{This
statement may not hold for exotic neutrino models
\cite{Beacom:2002vi,Beacom:2003eu} or for peculiar astrophysical
sources \cite{Serpico:2005sz,Kashti:2005qa,Serpico:2005bs}.}.
Despite the different behavior of the produced tau leptons with
respect to muons in terms of energy loss and decay length, both
$\nu_\mu$ and $\nu_{\tau}$ detection are sensitive to the matter
distribution near the NT site. Thus, a computation of the event
detection rate of a km$^3$ telescope requires a careful analysis of
the surroundings of the proposed site. The importance of the
elevation profile of the Earth surface around the detector was
already found of some relevance in Ref.\ \cite{Miele:2005bt}, where
some of the present authors calculated the aperture of the Pierre
Auger Observatory \cite{Auger,Abraham:2004dt} for Earth-skimming UHE
$\nu_{\tau}$'s. Indeed, air shower experiments can be used as NT's
at energies$\agt 10^{18}\,$eV, a topic recently reviewed in
\cite{Zas:2005zz}.  In particular, the possibility of detection of
the $\tau$ leptons produced by Earth-skimming UHE $\nu_\tau$'s has
been analyzed in a series of papers \cite{Miele:2005bt},
\cite{Capelle:1998zz}-\cite{Miele:2004ze}. In Ref.\
\cite{Miele:2005bt} the use of a Digital Elevation Map (DEM) of the
geographical area of the experiment proved useful to characterize
peculiar matter effects in Earth-skimming events.

The aim of this paper is to estimate the effective aperture for
$\nu_\tau$ and $\nu_\mu$ detection of a km$^3$ NT in the
Mediterranean sea placed at any of the three locations proposed by
the \verb"ANTARES", \verb"NEMO" and \verb"NESTOR" collaborations.
We do not consider any detail related to the experimental setup
nor the detector response. In particular, we assume full detection
efficiency via Cherenkov radiation for muons and taus crossing the
NT fiducial volume. This means that we consider a lepton as
detected if it crosses in any point the surface delimiting our
fiducial volume, without e.g. taking into account further
requirements or cuts needed for a good directional or energy
reconstruction. These would depend on several parameters, like the
spacing of the strings, the distribution of photomultipliers along
the string, etc. which are characteristics of the apparatus and
thus beyond the aim of the present analysis. We rather compare the
site characteristics by using the DEM of the different areas. We
shall therefore characterize and quantify the importance of
``matter effects'' for the three sites, and focus on the role
played by the geometry of the experiment in enhancing the effect.
These considerations may provide an important ingredient in
shaping the final design of a km$^3$ Mediterranean NT.

\begin{figure}[t]
\begin{center}
\hspace{2cm}
\includegraphics[width=.70\textwidth]{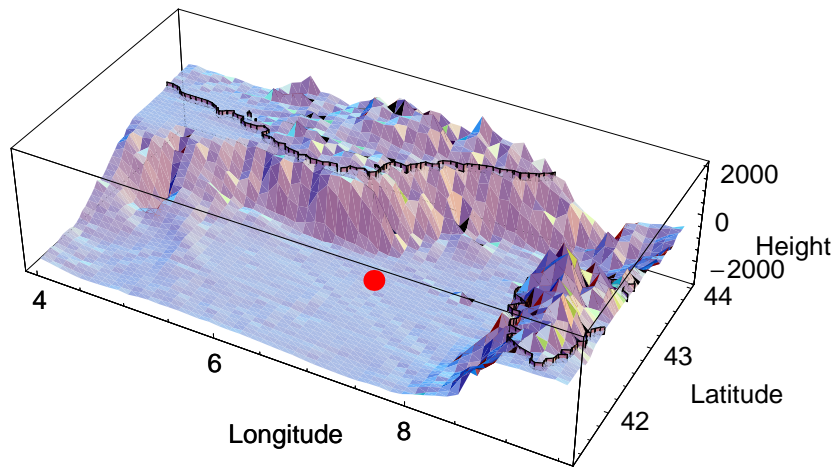}
\vspace{-0.0cm} \caption{The surface profile of the area near the
\texttt{ANTARES} site (red spot) at 42$^\circ$ 30' N, 07$^\circ$ 00'
E. The black curve represents the coast line. The sea plateau depth
in the simulation is assumed to be 2685 m. The effective volume
starts at an height of 100 m from the seabed, to account for the
spacing of the first photomultipliers as foresee by the current
designs. The km$^3$ detector is oriented along the E-W/S-N
directions.} \label{Antares}
\end{center}
\end{figure}

\begin{figure}[t]
\begin{center}
\hspace{2cm}
\includegraphics[width=.70\textwidth]{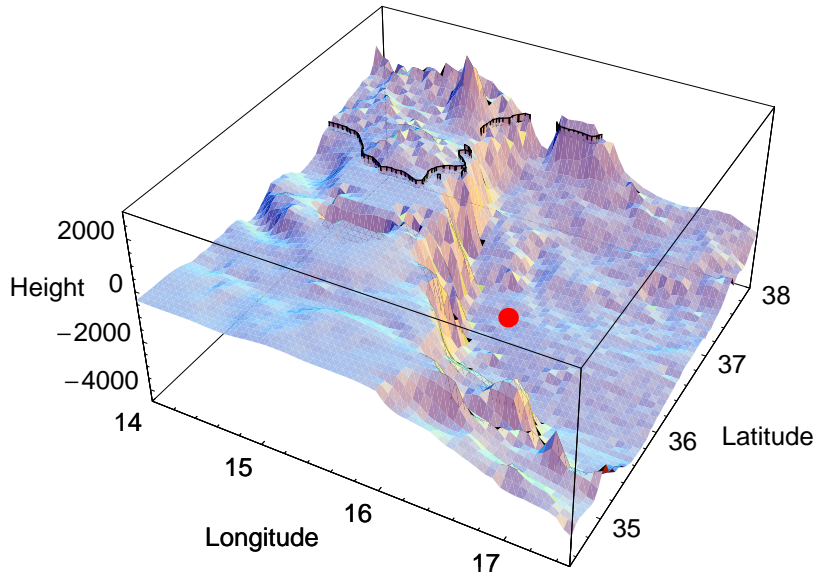}
\vspace{-0.8cm} \caption{The surface profile of the area near the
\texttt{NEMO} site (red spot) at 36$^\circ$ 21' N, 16$^\circ$ 10' E.
The black curve represents the coast line. The sea plateau depth
used in the simulation is 3424 m. The effective volume starts at an
height of 100 m from the seabed, to account for the spacing of the
first photomultipliers as foresee by the current designs. The km$^3$
detector is oriented along the E-W/S-N directions.} \label{Nemo}
\end{center}
\end{figure}

\begin{figure}[t]
\begin{center}
\hspace{2cm}
\includegraphics[width=.70\textwidth]{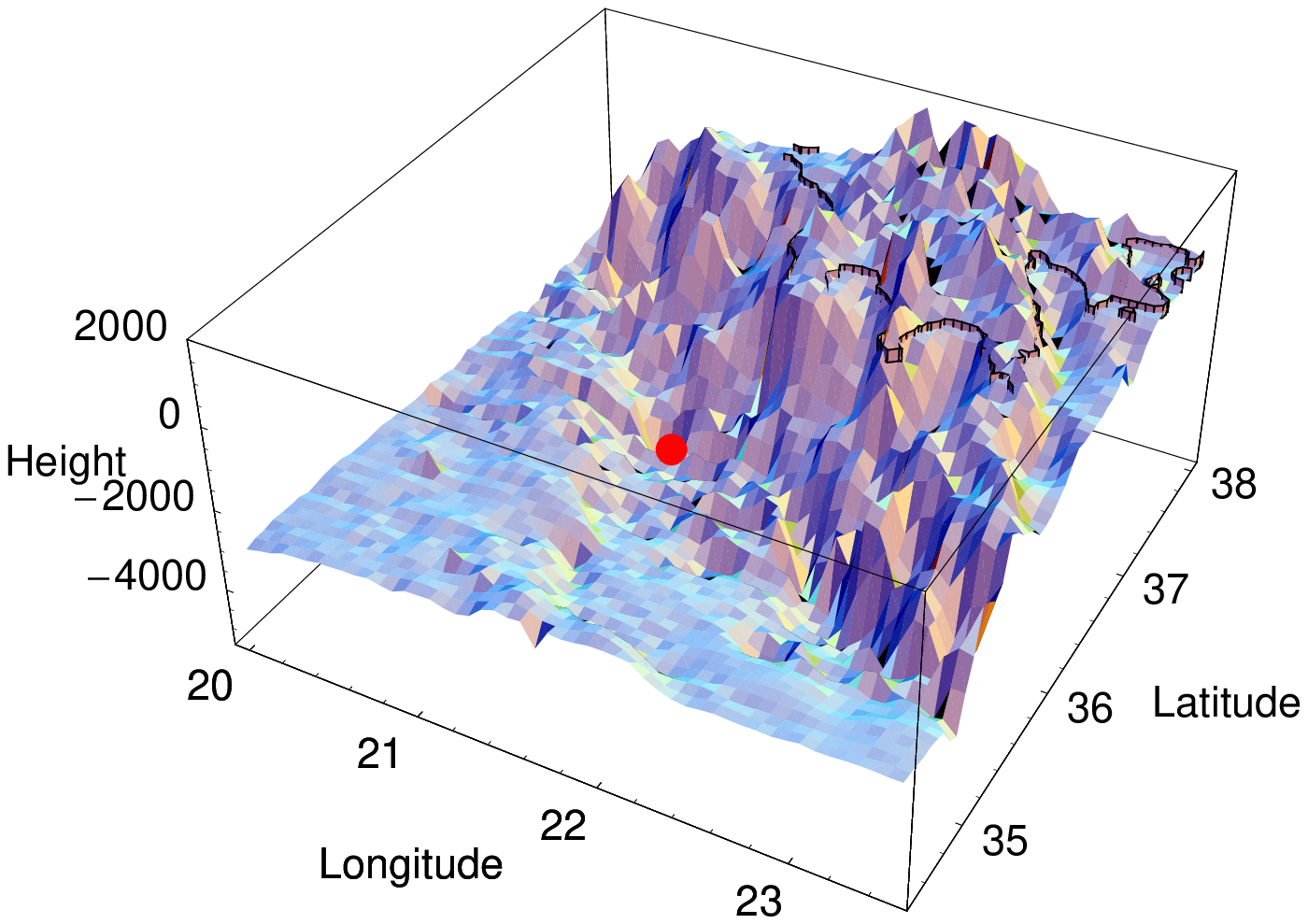}
\vspace{-0.8cm} \caption{The surface profile of the area near the
\texttt{NESTOR} site (red spot) at 36$^\circ$ 21' N, 21$^\circ$ 21'
E. The black curve represents the coast line. The sea plateau depth
in the simulation is assumed to be 4166 m. The effective volume
starts at an height of 100 m from the seabed, to account for the
spacing of the first photomultipliers as foresee by the current
designs. The km$^3$ detector is oriented along the E-W/S-N
directions.} \label{Nestor}
\end{center}
\end{figure}

A detailed DEM of the under-water Earth surface is available from
the Global Relief Data survey (ETOPO2) \cite{ETOPO2}, a grid of
altimetry measurements with a vertical resolution of 1 m averaged
over cells of 2 minutes of latitude and longitude. In Figures
\ref{Antares}, \ref{Nemo} and \ref{Nestor} we show the 3D maps of
the areas around the three NT sites. The black curve represents the
coast line, whereas the red spot stands for the location of the
apparatus. By following the same approach developed in
\cite{Miele:2005bt}, we use this DEM to produce a realistic and
statistically significant sample of $\nu_\tau$/$\tau$ and
$\nu_\mu$/$\mu$ tracks crossing the fiducial volume of the NT that
are then used to evaluate the effective aperture of each detector.

We note that when the events are reconstructed only in terms of
the energy loss along the track, the UHE taus can not be
distinguished from less energetic muons. This implies that the
reconstruction analyses of UHE $\nu_\mu$ and $\nu_\tau$ events are
highly entangled issues. We shall consider both of them, although
for the sake of clarity we shall first focus on $\nu_\tau$
detection. Of course, when considering shower events or more in
general contained events---where the charged lepton production
happens inside the instrumented volume---including events with
peculiar topologies like ``lollipop" or ``double bang", there are
realistic chances of flavor-tagging in the detector. However, in
the UHE range above $\sim 10^7$ GeV these kind of events are
subdominant with respect to the bulk of tau track-events. Further
details on the flavor discrimination possibilities are discussed
in Section \ref{nu_mu}.

The structure of the paper is as follows. In Section~\ref{ev_rate}
we introduce the formalism and definitions used in the analysis
and define the aperture for a NT. Our results for $\nu_\tau$
induced events are reported and discussed in Section~\ref{nu_tau}
for various incoming neutrino fluxes, while $\nu_\mu$/$\mu$ events
are described in Section~\ref{nu_mu}. Finally, we report our
conclusions in Section~\ref{concl}.

%%%%%%%%%%%%%%%%%%%%%%%%%%%%%%%%%%%%%%%%%%%%%%%%%%%%%%%%%%%%%%%%%%
\section{The effective aperture of a NT}\label{ev_rate}

\begin{figure}[t]
\begin{center}
\epsfig{figure=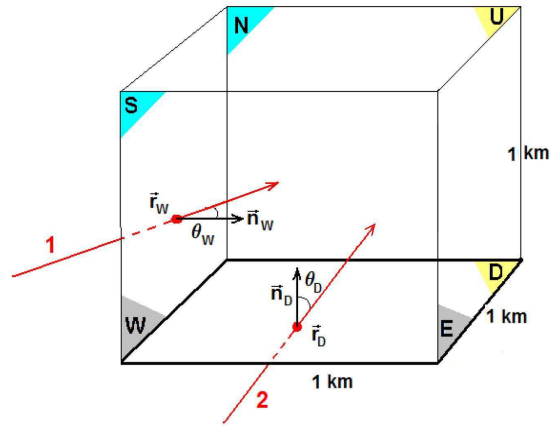,height=6cm} \caption{The angle
definition and the fiducial volume of a km$^3$ NT.} \label{kmcube}
\end{center}
\end{figure}

We define the km$^3$ NT {\it fiducial} volume as that bounded by
the six lateral surfaces $\Sigma_a$ (the subindex $a$=D, U, S, N,
W, and E labels each surface through its orientation: Down, Up,
South, North, West, and East), and indicate with $\Omega_a \equiv
(\theta_a, \phi_a)$ the generic direction of a track entering the
surface $\Sigma_a$. The scheme of the NT fiducial volume and two
examples of incoming tracks are shown in Fig.\ \ref{kmcube}. We
introduce all relevant quantities with reference to $\nu_\tau$
events, the case of $\nu_\mu$ being completely analogous.

Let $\d \Phi_\nu/(\d E_\nu \, \d\Omega_a)$ be the differential
flux of UHE $\nu_\tau + \bar{\nu}_\tau$. The number per unit time
of $\tau$ leptons emerging from the Earth surface and entering the
NT through $\Sigma_a$ with energy $E_\tau$ is given by
\bea \left( \frac{\d N_\tau}{\d t} \right)_a &=& \int \d\Omega_a \int
\d S_a \int \d E_\nu \, \frac{\d\Phi_\nu(E_\nu,
\Omega_a)}{\d E_\nu\,\d\Omega_a}\nonumber \\
& & \times\int
\d E_\tau \cos\left(\theta_a\right)
 k_a^\tau(E_\nu,E_\tau;\vec{r}_a,\Omega_a) \pp
\label{eq:1} \eea


This equation is the same as in \cite{Miele:2005bt}, but for full
duty cycle and detection efficiency.  The kernel
$k_a^\tau(E_\nu\,,E_\tau\,;\vec{r}_a,\Omega_a)$ is the probability
that an incoming $\nu_\tau$ crossing the Earth, with energy $E_\nu$
and direction $\Omega_a$, produces a $\tau$-lepton which enters the
NT fiducial volume through the lateral surface $\d S_a$ at the
position $\vec{r}_a$ with energy $E_\tau$ (see Fig.\ \ref{kmcube}
for the angle definition). If we split the possible events between
those with track intersecting the {\it rock} and the ones only
crossing {\it water}, the kernel
$k_a^\tau(E_\nu\,,E_\tau\,;\vec{r}_a,\Omega_a)$ is given by the sum
of these two mutually exclusive contributions,
\begin{equation}
k_a^\tau(E_\nu\,,E_\tau\,;\vec{r}_a,\Omega_a) =
k_a^{\tau,{r}}(E_\nu\,,E_\tau\,;\vec{r}_a,\Omega_a)+
k_a^{\tau,{w}}(E_\nu\,,E_\tau\,;\vec{r}_a,\Omega_a) \pp
\label{kern-split}
\end{equation}
Let us focus on the {\it rock} events contributing to
$k_a^{\tau,{r}}(E_\nu\,,E_\tau\,;\vec{r}_a,\Omega_a)$. These can
be classified according to their production mechanism as follows:
\\
1) events in which the $\nu_\tau$ interacts producing a $\tau$ in
the rock (r1);\\
2) events in which the $\nu_\tau$ interacts producing a $\tau$ in
water, on the way to the NT (r2);\\
3) events in which the $\nu_\tau$ interacts producing a $\tau$ inside
the NT fiducial volume (r3). \\
Therefore one has
\begin{eqnarray}
k_a^{\tau,{r}}(E_\nu\,,E_\tau\,;\vec{r}_a,\Omega_a)&=&
k_a^{\tau,{r1}}(E_\nu\,,E_\tau\,;\vec{r}_a,\Omega_a)+
k_a^{\tau,{r2}}(E_\nu\,,E_\tau\,;\vec{r}_a,\Omega_a)\nonumber\\
&+& k_a^{\tau,{r3}}(E_\nu\,,E_\tau\,;\vec{r}_a,\Omega_a)\pp
\label{kernrock-split}
\end{eqnarray}
Although here, for the sake of brevity, we only discuss in details
the events occurring in rock (r1), the analysis of those of type
(r2) and (r3) is completely analogous and straightforward. Of
course, all contributions (r1)-(r3) have been added to compute the
event rate.

As already shown in details in \cite{Miele:2005bt,Aramo:2004pr} a
(r1)-event corresponds to the simultaneous fulfillment of the
following conditions:
\begin{itemize}
\item[1)] A $\nu_\tau$ with energy $E_\nu$ travels over a distance $z$
through the Earth before interacting. The corresponding probability
$P_1$ is given by
\begin{equation}
P_1=\exp\left\{-\frac{z}{\lambda_{CC}^\nu(E_\nu)}\right\} \vv
\label{eq:2} \ee where \be \lambda_{CC}^\nu(E_\nu) =
\frac{1}{\sigma_{CC}^{\nu N}(E_\nu)\,\varrho_r\,N_A} \vv
\end{equation}
where $N_A$ the Avogadro number. See
\cite{Miele:2005bt,Aramo:2004pr} for notations as well as a detailed
discussion of the neutrino-nucleon cross section, $\sigma_{CC}^{\nu
N}(E_\nu)$. In the present formalism the effect of the Earth density
profile is approximated using, track by track, the averaged
$\varrho_r$ along the chord subtended by that track. The
calculations are made with the parametrization of the Earth density
profile of Ref. \cite{EarthModel}. Note, however, that for almost
horizontal events particles travel in the terrestrial crust only,
and thus the superficial value for the Earth density $\varrho_r
\simeq 2.65$ g/cm$^{3}$ \cite{Feng:2001ue} would be an accurate
approximation. Some differences could appear for low energy
particles deeply crossing the Earth. We checked that, using the
constant value of the crust density for all the Earth density
profile the changes in the tau aperture are generally less than
10\%, while the effect is of the order of 16\% for muons with E
$<10^5$ GeV, i.e. below the energy range for which the cosmic flux
is expected to dominate over the atmospheric flux. The inclusion of
the Earth density profile also affect at the 10\% level the
distributions of both $\tau$s and $\mu$s incoming zenith angles: due
to increased screening along the nadir direction (vertical
up-going), the distributions slightly shrink along the horizontal
direction.

\item[2)] The neutrino produces a $\tau$ in the interval $z,
z+\d z$, the probability of such an event being
\begin{equation}
P_2 \, \d z= \frac{\d z}{\lambda_{CC}^\nu\,(E_\nu)} \pp \label{eq:4}
\end{equation}
We do not consider here the event corresponding to the scattering
of a $\nu_\tau$ via neutral current in the Earth followed by
conversion via charged current, which amounts to a small
distortion of the incoming neutrino flux, the latter being yet
unknown \cite{L'Abbate:2004hv}. Of course, this sub-leading effect
should be added when trying to reconstruct the flux from
experimental data. Also, we consider the charged lepton track as
collinear with the parent neutrino direction, which is highly
accurate given the huge relativistic boosting factors involved.
\item[3)] The produced $\tau$ emerges from the Earth rock with
an energy $E_\tau'$. This happens with a probability
\begin{eqnarray}
P_3 &=& \exp\left\{-\frac{m_\tau}{c \tau_\tau \beta_\tau
\varrho_r} \left(\frac{1}{E_{\tau}'}-
\frac{1}{E_\tau^0(E_\nu)}\right)\right\} \nonumber \\
& & \times \delta\left( E_{\tau}'-E_\tau^0(E_\nu) \, e^{-
\beta_\tau \varrho_r (z_{r} -z)}\right) \vv \label{eq:9}
\end{eqnarray}
\noindent where $m_\tau = 1.78$ GeV, $\tau_\tau \simeq 3.4 \times
10^{-13}\,$s is the $\tau$ mean lifetime and $E_\tau^0$ is the
$\tau$ energy at production, whereas the parameter $\beta_\tau =
0.71 \times 10^{-6}$ cm$^2$ g$^{-1}$ weights the leading term in the
$\tau$ differential energy loss in rock
\cite{Aramo:2004pr,Dutta:2005yt}, \be \frac{\d E_\tau}{\d z}=-
\left( \beta_\tau + \gamma_\tau E_\tau \right) E_\tau \varrho_r \pp
\label{dedz} \ee The contribution of $\gamma_\tau$ can be neglected
as it only affects extremely energetic $\tau$'s which, differently
from the case of the Pierre Auger Observatory, are not relevant
for NT's. The quantity $z_{r}\left(\vec{r}_a\,,\Omega_a\right)$
represents the total length in rock for a given track entering the
lateral surface $\Sigma_a$ of the fiducial volume at the point
$\vec{r}_a$ and with direction $\Omega_a$.

\item[4)] Finally, the $\tau$ lepton emerging from the Earth
rock propagates in water and enters the NT fiducial volume
through the lateral surface $\Sigma_a$ at the point $\vec{r}_a$
with energy $E_\tau$. The corresponding survival probability is
\begin{eqnarray}
P_4 &=& \exp\left\{-\frac{m_\tau}{c \tau_\tau \beta_\tau
\varrho_w} \left(\frac{1}{E_{\tau}}-
\frac{1}{E_{\tau}'}\right)\right\}\, \delta\left(
E_{\tau}-E_{\tau}' \, e^{- \beta_\tau \rho_w z_{w}
}\right) \vv \nonumber\\\label{eq:10}
\end{eqnarray}
where $\varrho_w$ stands for the water density and
$z_{w}\left(\vec{r}_a\,,\Omega_a\right)$ represents the total
length in water before arriving to the fiducial volume for a given
track entering the lateral surface $\Sigma_a$ at the point
$\vec{r}_a$ and with direction $\Omega_a$.
\end{itemize}

Collecting together the different probabilities in Eqs.
(\ref{eq:2}), (\ref{eq:4}), (\ref{eq:9}) and (\ref{eq:10}), we
have
\begin{equation}
k_a^{\tau,{r1}}(E_\nu\,,E_\tau\,;\vec{r}_a,\Omega_a)=\int_0^{z_r}\d z\,\int_0^{E_\tau^0(E_\nu)}
\d E_{\tau}'\, \,P_1 \, P_2 \, P_3 \, P_4 \pp \label{eq:12}
\end{equation}
Similar results can be obtained for the (r2)- and (r3)-events as
well as for those we defined as {\it water}-like.

For an isotropic flux we can rewrite Eq. (\ref{eq:1}), summing
over all the surfaces, as
\begin{eqnarray}
\frac{\d N_\tau^{(r,w)}}{\d t} &=& \int \d E_\nu \,
\,\frac{1}{4\pi}\,\frac{\d\Phi_\nu(E_\nu)}{\d E_\nu}
\,A^{\tau(r,w)}(E_\nu)= \nonumber \\ &=& \sum_a \int \d E_\nu \,
\,\frac{1}{4\pi}\,\frac{\d\Phi_\nu(E_\nu)}{\d E_\nu}
\,A_a^{\tau(r,w)}(E_\nu) \vv \label{kernel1}
\end{eqnarray}
which defines the total aperture $A^{\tau(r,w)}(E_\nu)$, with ``$r$"
and ``$w$" denoting the {\it rock} and {\it water} kind of events,
respectively. The contribution of each surface to the total aperture
reads
\begin{eqnarray}
A_a^{\tau(r,w)}(E_\nu) &=& \int \d E_\tau \int \d\Omega_a \int  \d S_a
\, \cos\left(\theta_a\right) \,
k_a^{\tau,{(r,w)}}(E_\nu\,,E_\tau\,;\vec{r}_a,\Omega_a) \pp
\label{kernel2}
\end{eqnarray}

\section{The event rate for $\nu_\tau$ interactions}\label{nu_tau}

\begin{figure}[t]
\begin{center}
\epsfig{figure=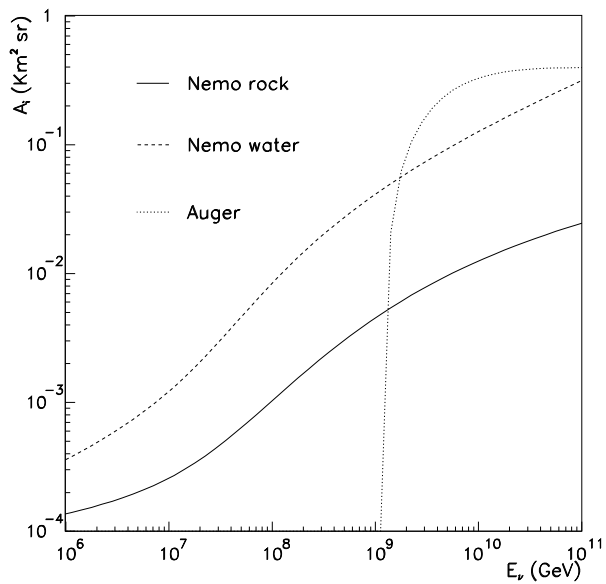,height=8cm} \caption{The effective
apertures $A^{\tau(r)}(E_\nu)$ (solid line) and $A^{\tau(w)}(E_\nu)$
(dashed line) defined in Eq.\ (\ref{kernel1}) versus tau neutrino
energy for \texttt{NEMO}. The dotted line corresponds to the same
quantity for the Auger Fluorescence Detector for Earth-skimming
$\nu_\tau$ as in \cite{Miele:2005bt}.} \label{apertures}
\end{center}
\end{figure}
We show in Fig. \ref{apertures} the apertures $A^{\tau(r,w)}$ for
the \verb"NEMO" site together with the corresponding quantity for
the Pierre Auger Observatory Fluorescence Detector (FD) calculated
in \cite{Miele:2005bt}. Note that the Auger case is only for
Earth-skimming $\tau$'s, since down-going neutrino induced events
can be disentangled from ordinary cosmic rays only for very inclined
showers.  Interestingly, the \verb"NEMO"-{\it water} and Auger-FD
apertures almost match at the FD threshold of $10^{18}\,$eV, so that
using both detectors results into a wide energy range of sensitivity
to $\nu_\tau$ fluxes.
\begin{figure}[!tb]
\begin{center}
\begin{tabular}{cc}
\epsfig{figure=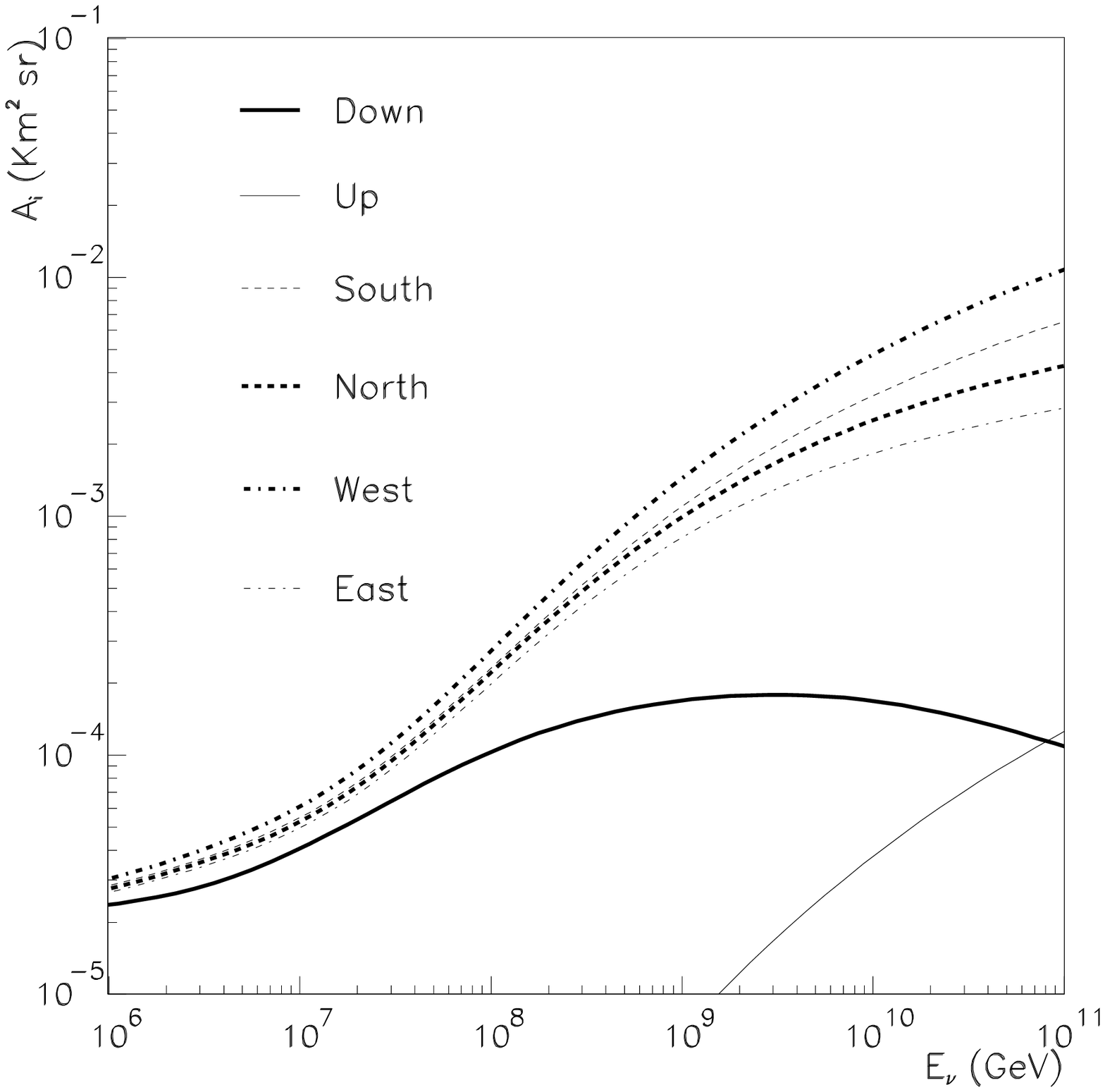,height=7cm} &
\epsfig{figure=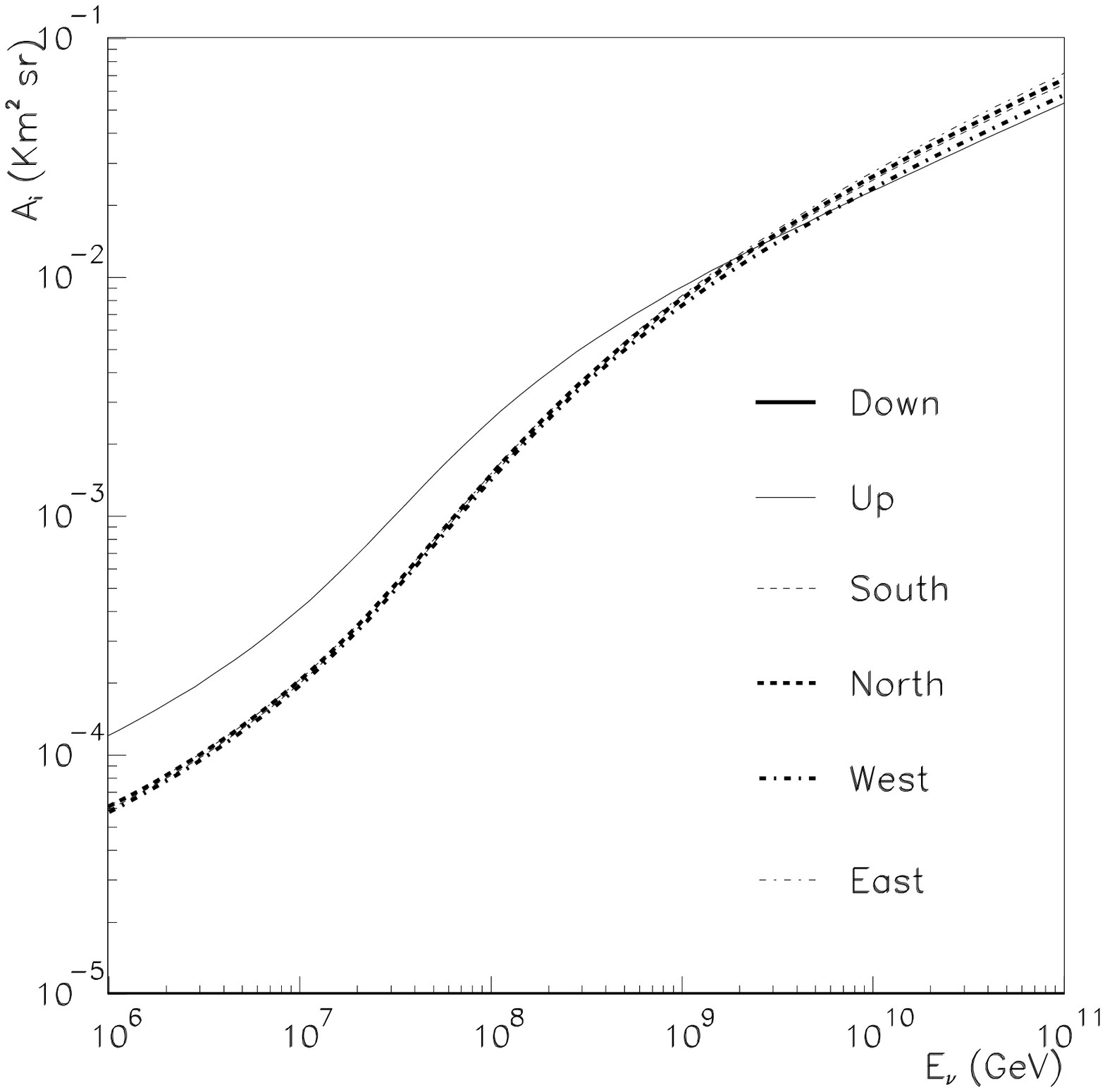,height=7cm}
\end{tabular}
\end{center}
\caption{The effective apertures $A_a^{\tau(r,w)}(E_\nu)$ of Eq.\
(\ref{kernel2}) versus tau neutrino energy for (left) {\it rock}
events and (right) {\it water} events for the \texttt{NEMO} site.}
\label{apertsurf}
\end{figure}
\begin{figure}[t]
\begin{center}
\epsfig{figure=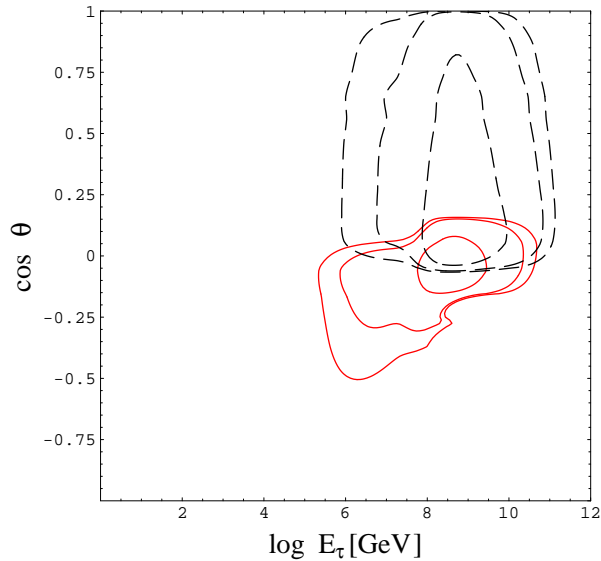,height=8cm} \caption{ Contour plot in
the plane zenith angle-$\tau$ energy for the \texttt{NEMO} site and
for {\it rock} (red full lines) and {\it water} (black dashed lines)
events. In both cases the contours enclose 68, 95 and 99 \% of the
total number of events calculated assuming a GZK-WB flux
(\cite{Waxman:1998yy}, see also \cite{Miele:2004ze}). $\cos \theta =
1, 0, -1$ correspond respectively to down-going, earth-skimming, and
up-going events. } \label{nemocont}
\end{center}
\end{figure}

We show in Fig.\ \ref{apertsurf} the high energy behavior for each
surface contributing to the effective aperture. For {\it rock}
events there is a clear W-E asymmetry, easily understood in terms of
matter effects related to the particular morphology of the
\verb"NEMO" site (see Fig.\ \ref{Nemo}). A much smaller S-N
asymmetry is also present. In other words, the asymmetries in the
number of rock-events reflect the asymmetries in the morphology of
the site.

For neutrino energies larger than $10^7$ GeV the main contribution
to the aperture $A^{\tau(r)}(E_\nu)$ comes from the lateral
surfaces, i.e.\ from $\tau$ leptons emerging from the rock far
from the NT basis and crossing the fiducial volume almost
horizontally. Instead, the upper surface contribution is
negligible due to the very small fraction of events crossing the
rock and entering the detector from above. The decreasing
contribution of the bottom face to rock events is due to the Earth
shadowing effect.

For {\it water} events the contribution to the aperture from all
surfaces is comparable (except for the lower one which has no
events), the upper one providing a slightly larger contribution as
the energy decreases. Indeed, events which would cross the lateral
surfaces should travel over a longer path in water and this becomes
more unlikely at lower energies due to the shorter $\tau$ decay
length. The matter effect in the case of {\it water} events is less
pronounced and  anticorrelated with the asymmetries in the
morphology of the site resulting effectively in a small (percent)
screening effect.

In Fig.\ \ref{nemocont} we report, for both the {\it rock} and {\it
water} cases and for the \texttt{NEMO} site, the contours enclosing
68, 95 and 99 \% of the total event rate, as they appear in the
plane $E_\tau$-$\theta$ plane, where $\theta$ is the arrival
direction zenith angle. These results were obtained assuming a
Waxmann-Bahcall-like neutrino flux (GZK-WB) \cite{Waxman:1998yy}
(see also \cite{Miele:2005bt} and references therein). As the energy
increases, the arrival directions of {\it rock} events are almost
restricted to the horizontal (Earth-skimming), while at lower
energies the earth-screening effect is less pronounced and this
explains the broader angular distribution. The situation is
different for {\it water} events for which the angular distribution
is broad at all energies, a purely geometrical effect due to the
fact that (down-going) {\it water} events are not screened in few
kilometers of water. The same geometrical considerations explain the
ratio of {\it water} to {\it rock} event rate of $\cal{O}$(10) (see
Table \ref{table::events-WB}) which is simply related to the ratio
of down-going to Earth-skimming solid angles. This is the same kind
of behavior expected in the Auger detector although, as we already
mentioned, the down-going events in this case are hardly
distinguishable from the background of proton-induced showers so
that only Earth-skimming or almost horizontal showers can be used to
identify unambiguously the neutrino-induced events. It is also worth
commenting the expected $\tau$ energy distribution shown in Fig.\
\ref{nemocont}.  All events correspond to a relatively narrow energy
window, from $10^6$ GeV up to $10^{10}$ GeV, where the lower cut-off
arises from the shorter $\tau$ decay length at low energy.
\begin{figure}[t]
\begin{center}
\epsfig{figure=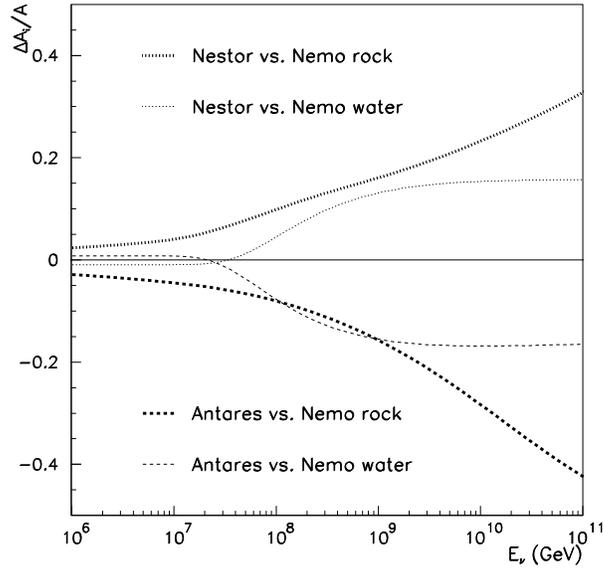,height=8cm}
\caption{A comparison of the effective apertures
$A^{\tau((r,w))}(E_\nu)$ for the three NT sites. We plot the
ratios
$[A^{\tau((r,w))}(\mbox{\texttt{NESTOR}})-A^{\tau((r,w))}(\mbox{\texttt{NEMO}})]/A^{\tau((r,w))}(\mbox{\texttt{NEMO}})$
and
$[A^{\tau((r,w))}(\mbox{\texttt{ANTARES}})-A^{\tau((r,w))}(\mbox{\texttt{NEMO}})]/A^{\tau((r,w))}(\mbox{\texttt{NEMO}})$
versus the neutrino energy.}
\label{comptelesc}
\end{center}
\end{figure}

In Fig.\ \ref{comptelesc} we compare the detection performances of
a km$^3$ NT placed at one of the three sites in the Mediterranean
sea. The \verb"NESTOR" site shows the highest values of the
$\tau$-aperture for both {\it rock} and {\it water}, due to its
larger depth and the particular matter distribution of the
surrounding area, while the lowest rates are obtained for
\verb"ANTARES". The aperture in the three sites can be quite
different at high energy, but in order to get the expected number
of UHE events per year, one has to convolve the aperture with a
neutrino flux which typically drops rapidly with the energy.
Although the percentage value of the matter effects remains
unchanged, in this very low statistics regime they can be hardly
distinguished; still, they can be enhanced by an appropriate
choice of the detector shape as we discuss in the following.
\begin{table}[t]
\centering
\begin{tabular}{|c|c|c|c|}
\hline Surf.
 & \verb"ANTARES" & \verb"NEMO" & \verb"NESTOR"\\
\hline
D & 0.0059/0 & 0.0059/0 & 0.0058/0 \\
U & 0/0.1677 & 0.0002/0.2133 & 0.0002/0.2543 \\
S & 0.0185/0.1602 & 0.0256/0.1773 & 0.0240/0.2011 \\
N &  0.0241/0.1540& 0.0229/0.1823 & 0.0321/0.1924 \\
W & 0.0212/0.1584 & 0.0335/0.1691 & 0.0265/0.2002 \\
E & 0.0206/0.1589 & 0.0190/0.1875 & 0.0348/0.1907 \\
\hline Total & 0.090/0.799 & 0.107/0.929 & 0.123/1.039\\
\hline
\end{tabular}
\caption{Estimated rate per year of {\it rock/water} $\tau$ events
at the three km$^3$ NT sites for a GZK-WB flux
(\cite{Waxman:1998yy,Miele:2004ze}). The contribution of each
detector surface to the total number of events is also reported.}
\label{table::events-WB}
\end{table}

Knowing the aperture of the NT at each site, we can compute the
expected $\tau$ event rate, once a neutrino flux is specified. In
Table \ref{table::events-WB} these rates are shown assuming a
GZK-WB flux (\cite{Waxman:1998yy,Miele:2004ze}). The effect due to
the local matter distribution is responsible for the N-S, W-E and
NE-SW asymmetries for the \verb"ANTARES", \verb"NEMO" and
\verb"NESTOR" sites, respectively, as expected from the matter
profiles shown in Figs.\ \ref{Antares}, \ref{Nemo} and
\ref{Nestor}. These matter effects, for the specific UHE flux
considered (GZK-WB), correspond to an enhancement of {\it rock}
events which goes from 20 to 50\% for the three sites,
respectively, and a screening factor for {\it water} events from 3
to 10\%. The largest relative difference among lateral surfaces is
in the case of W/E for \verb"NEMO", where the huge wall to the
west of the site (see Fig. \ref{Nemo}) improves the rate by about
75\%, almost a factor 2! Notice also that the {\it water} events
from the U surface are basically proportional to the depth.

It is important to emphasize that the impact of matter effects on
the rates depends critically upon the energy spectrum of the UHE
neutrino flux. For more energetic neutrino fluxes the enhancement
factor is expected to be more significant (see the energy behavior
of $A_a^{\tau(r)}(E_\nu)$ in Fig.\ \ref{apertsurf}). In Table
\ref{table::events} the rate of {\it rock/water} $\tau$ events are
computed for the three different km$^3$ NT sites using several UHE
neutrino fluxes as already considered in
\cite{Miele:2005bt,Aramo:2004pr} and described in
\cite{Waxman:1998yy}-\cite{Bhattacharjee:1998qc} (see also Fig.\
\ref{sensitivities}). For comparison, we also show in the last
column the corresponding prediction for Earth-skimming $\nu_\tau$
at Auger-FD. As can be seen from Table \ref{table::events}, the
relative enhancements due to matter effects on {\it rock} events
can be as large as 30\%, whereas the difference in the rates of
{\it water} events for a fixed neutrino flux is mainly due to the
different depth of the three sites.

An interesting feature is the dependence of the event rate upon
the shape of the NT detector for a fixed total volume of 1 km$^3$,
a property that might be relevant for the eventual design of the
detector. Consider for example a km$^3$ NT placed at the
\verb"NEMO" site with the shape of a parallelepiped rather than a
cube, where in particular the E and W surfaces are enlarged by a
factor 3 in the horizontal dimension, the N and S surfaces being
reduced by the same factor, keeping the height of towers still of
1 km. In this case the expected rate of {\it rock} events per year
is enhanced by almost a factor 2, from 0.11 to $0.18$ for the
GZK-WB flux, while this enhancement could be even larger for
neutrino fluxes with a larger high energy component. Moreover, the
expected rate of {\it water} events increases as well by a factor
of the order of 50\%, from $0.93$ up to $1.40$ per year. Similar
exercises can be also performed for the \verb"ANTARES" and
\verb"NESTOR" sites. Of course, a further possibility which might
favor UHE-$\tau$ detection consists in increasing the effective
volume of the detector keeping unchanged the 1 km height and the
number of towers of photomultipliers but adopting a larger
spacing. As an example, for a factor four larger volume with a
doubling of the tower spacing both the {\it rock} and {\it water}
$\tau$ events would increase by almost a factor two, but obviously
at the expense of the energy threshold and the quality of the
event reconstruction for ``low-energy'' (TeV) neutrinos.  For a
detector aiming at the exploration of the range above the PeV,
this is a less severe problem.
\begin{table}[t] \centering
\begin{tabular}{|c|c|c|c|c|}
\hline $\nu$-fluxes
 & \verb"ANTARES" & \verb"NEMO" & \verb"NESTOR" & Auger-FD \cite{Miele:2005bt}\\
\hline
GZK-WB& 0.090/0.799 & 0.107/0.929 & 0.123/1.039 & 0.074\\
GZK-L & 0.099/1.076 & 0.130/1.282 & 0.157/1.465 & 0.213\\
GZK-H & 0.225/2.744 & 0.313/3.280 & 0.386/3.766 & 0.560\\
NH    & 0.891/8.696 & 1.102/10.19 & 1.295/11.47 & 1.245\\
TD    & 0.701/5.072 & 0.817/5.799 & 0.921/6.424 & 0.548\\
\hline
\end{tabular}
\caption{Yearly rate of {\it rock/water} $\tau$ events at the
three km$^3$ NT sites for different UHE neutrino fluxes.  GZK-H is
for an initial proton flux $\propto 1/E$, assuming that the EGRET
flux is entirely due to $\pi$- photoproduction.  GZK-L shows the
neutrino flux when the associated photons contribute only up to
20\% in the EGRET flux. GZK-WB stands for an initial proton flux
$\propto 1/E^2$ \cite{Waxman:1998yy}--\cite{Semikoz:2003wv}. The
other two neutrino fluxes correspond to more exotic UHECR models.
NH represents the neutrino flux prediction in a model with new
hadrons \cite{Kachelriess:2003yy}, whereas TD is the neutrino flux
for a topological defect model \cite{Bhattacharjee:1998qc}. In the
last column we report the corresponding prediction for
Earth-skimming $\nu_\tau$ at Auger-FD.} \label{table::events}
\end{table}

The fact that the event rate depends upon the total surface of the
detector is a peculiar feature of a NT, quite differently from what
expected at the Auger observatory. Actually, in this case observed
showers are generally initiated not very far from the detector
compared to its dimensions so that the shape of the detector (i.e.,
the position on the border where the FDs are placed) is not as
important as its ``volume'' (controlled by the area enclosed by the
FDs). In fact, in order to produce a $\tau$ emerging from the Earth
with enough energy to generate a shower detectable by the Auger FDs,
the energy of the neutrino should be larger than 1 EeV$=10^{18}$ eV,
taking into account the $\tau$ energy loss in the rock.  But the
decay length of such a UHE $\tau$ is $l_{\rm decay} \simeq 50\,$km
$\times (E_\tau/{\rm EeV})$, to be compared with the dimensions of
the Auger fiducial volume, $\sim 50\times 60\times10$ km$^3$.
Conversely, a neutrino telescope can detect taus or muons which are
produced very far from the detector by a neutrino charged-current
interaction, from distances comparable to the charged lepton range
at that particular energy \cite{Yoshida:2003js}. Indeed, the $\tau$
range in water is of the order of several kilometers: from the value
of $\beta_{\tau}=0.71 \times 10^{-6}$ cm$^2$ g$^{-1}$ we obtain an
attenuation length $1/(\beta_{\tau}\varrho_{w}) \simeq 15$ km, while
for muons (see next section) the range is approximately eight times
smaller, of the order of 2 km. In other words, the effective volume
of a NT of the kind discussed so far can be much larger than 1
km$^3$, thus maximizing the detector area might greatly improve the
detection rate.

Of course, one should not forget that the design of a NT also
depends strongly upon more detailed experimental considerations.
Shapes which are not very compact or a detector with very sparse
instrumentation have worse performances in the reconstruction of
track properties as well as in signal-background separation, though
this is mainly problematic at energies lower than 100 TeV, in the
atmospheric neutrino energy range. In any case, our analysis
suggests that the choice of the detector shape could be an important
feature in orienting the target of a NT investigation towards either
atmospheric or extra-atmospheric neutrino physics. In this respect,
the possibility to modify this parameter quite easily for a NT water
detector offers a great advantage with respect to an under-ice
detector.

A comment is in turn regarding the various approximations we used in
the calculation. In particular we neglected tau regeneration
effects. It is well known that these effects depend on the adopted
incoming spectrum with a typical behavior in which the steeper the
spectrum the less relevant the effect is
\cite{Halzen:1998be,Bugaev:2003sw}. For an $E^{-2}$ spectrum the
effect is almost negligible \cite{Halzen:1998be}, while for harder
spectra like the ones we presently consider the effect can be of the
order of 20\% for taus coming from the nadir direction and of
decreasing relevance for more horizontal events
\cite{Bugaev:2003sw}. An estimate of the direction averaged effect
gives then a correction less of 10\%.

A further approximation regards the stochastic nature of the tau
interaction in matter that we approximated like a continuous energy
loss process through the parametrization of Eq.\ (\ref{dedz}). At
energies larger than 10$^6$ GeV, tau energy losses are affected by
the large theoretical uncertainty on cross-section for photonuclear
interaction, the leading mechanism at these energies (see
\cite{Aramo:2004pr} and \cite{DeYoung:2006fg}, and references
therein). Differently from muons, for taus the dominant source of
uncertainty is not the stochastic vs. continuous nature of the
energy loss, but the model-dependence of the photonuclear
interaction itself; the stochastic nature of the losses is then
sub-dominant with respect to the understanding of the process. A
detailed discussion of the problem is given in Ref.
\cite{Aramo:2004pr}. The continuous approximation is then enough for
the estimate of the mean rate values as given in the text,
especially for a relative comparison of the sites. A more careful
treatment would be needed if one is interested in a realistic
estimate of the errors.

A final comment deserves the dependence that matter effects could
have not only on truly differences of amount of matter present in
different directions, but also on the differences in the lepton
interactions in water or rock due to the different chemical
composition $(A,Z)$, i.e. to differences between $\beta_{r}$ and
$\beta_w$. We study the issue  performing detailed calculations of
the lepton propagation as given in \cite{Bottai:2000en}. The
calculations show that the (A,Z) dependence is of the order 10\% for
taus, almost constant at high energies ($>1$ PeV), while of the
order 20\% for muons again almost constant at high energies. Given
the model uncertainties in the tau losses and the level of
approximation of 10\% used throughout the paper we used the same
value of $\beta$ for rock and water losses for both muons and taus
so that the differences seen at high energies in the apertures  in
Fig.\ref{apertsurf} has to be all ascribed to a genuine matter
effect. Note moreover that the matter effect is a feature increasing
with energy, amounting for example in a factor even of four in
difference between the $W-E$ surfaces for \verb"NEMO" at $10^{11}$
GeV, while the percentage difference in $\beta_w/\beta_r$ always
remain of the order 10\% through the whole energy range. The role of
chemical composition then is at most subdominant.

\begin{figure}[!tb]
\begin{center}
\begin{tabular}{cc}
\epsfig{file=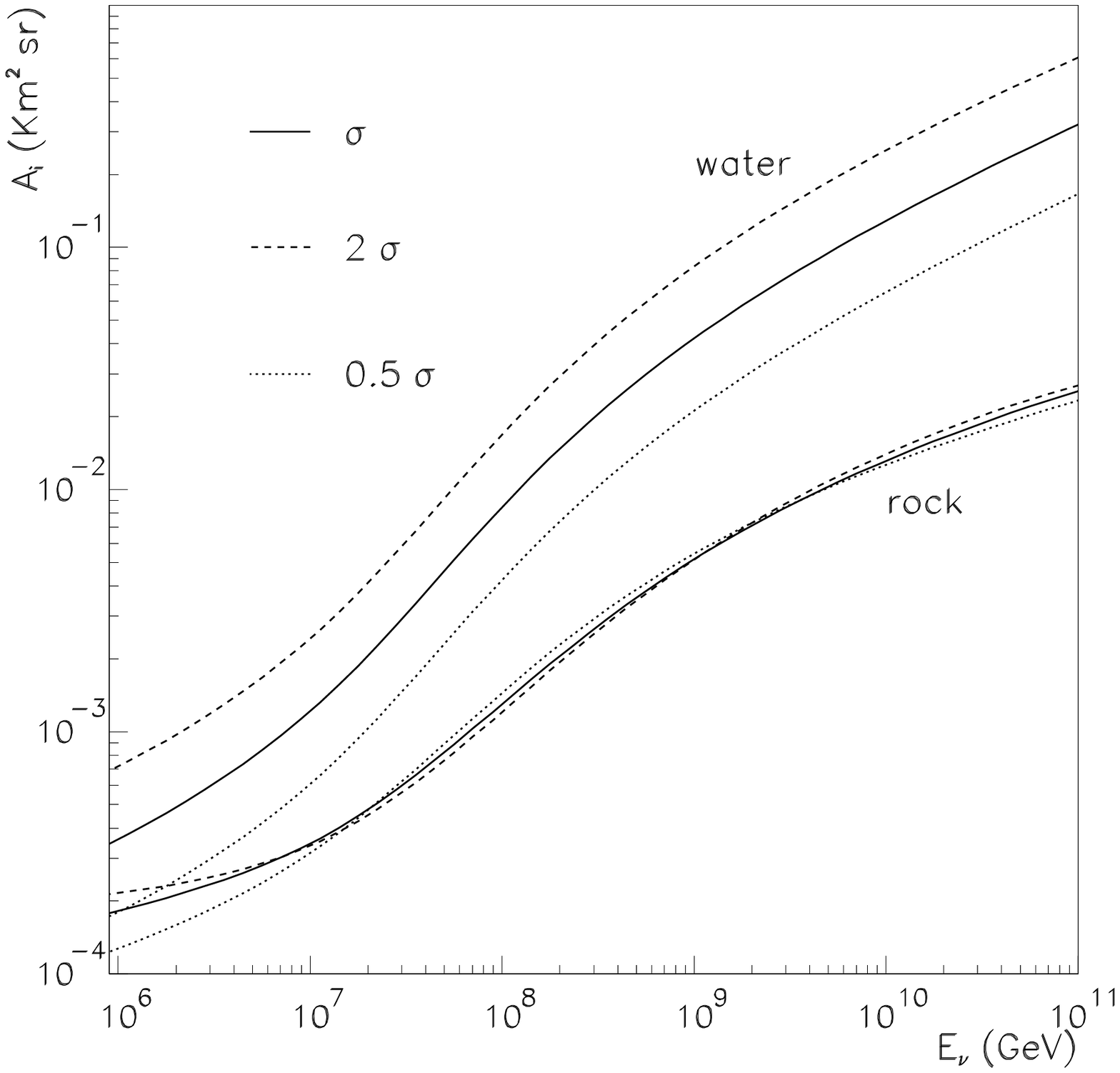,height=7cm} &
\epsfig{file=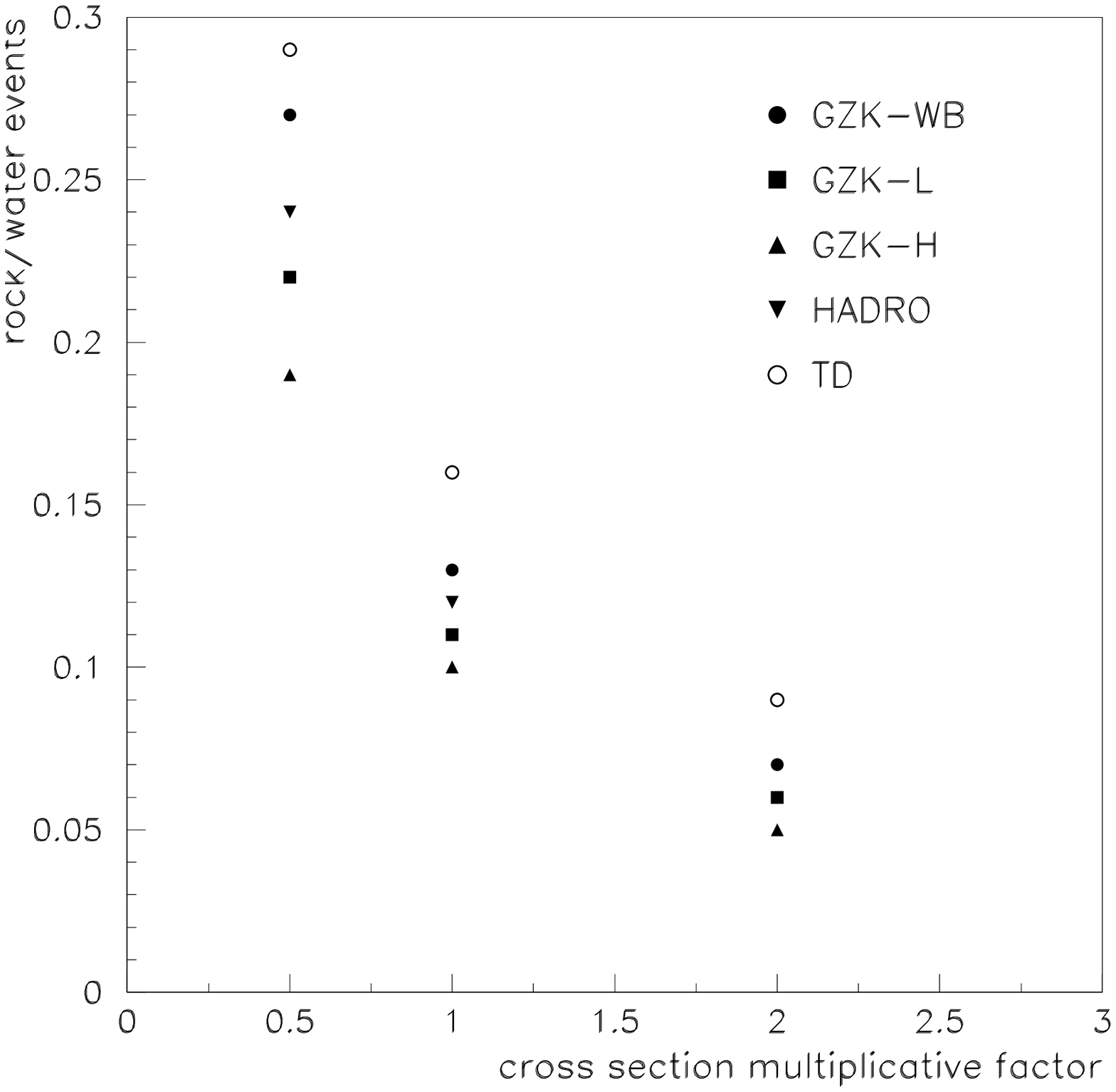,height=7cm}
\end{tabular}
\end{center}
\caption{(Left) The effective apertures $A^{\tau(r,w)}(E_\nu)$ for the
\texttt{NEMO} site for a $\nu_\tau$--nucleon cross section multiplied
by a factor 0.5, 1 and 2 with respect to the standard result
$\sigma$. (Right) Ratios of the number of events in {\it rock/water}
when the cross section is rescaled by the factor shown on the $x$
axis, for several incoming UHE neutrino fluxes.} \label{AperEventsig}
\end{figure}

One of the main motivations for studying UHE neutrinos is that they
provide a possibility to explore a range of energies for scattering
processes which is still untested (maybe impossible to test) by
particle accelerators. In this respect, measuring the
neutrino--nucleon cross sections at high energies could have a large
impact on constraining or discovering new physics beyond the
standard model (see e.g.\
\cite{Anchordoqui:2005is,Anchordoqui:2005gj}).  While a measurement
of the event energy spectrum cannot remove in general a degeneracy
between the neutrino cross section and the incoming neutrino flux, a
neutrino telescope could offer the interesting capability of
disentangling these two factors because of the role of matter
effects. Indeed, provided that enough statistics is collected and
the detector has a good zenith angle resolution, the flux dependence
can be subtracted off by measuring the ratio of the event rates
coming from different directions
\cite{Kusenko:2001gj,Hooper:2002yq,Hussain:2006wg}. In the left
panel of Fig.\ \ref{AperEventsig} we show how the \verb"NEMO"
effective apertures for $\tau$ {\it rock} and {\it water} events
change if the neutrino-nucleon cross section is half or twice the
standard model result, while in the right panel we display the ratio
of {\it water/rock} $\tau$ event rates for several adopted fluxes.
We see that this ratio is quite sensitive to the value of the cross
section. In particular, the number of {\it rock} events is
essentially unaffected while the {\it water} event rate increases
almost linearly with the cross section. Clearly, since the
statistical error on the ratio would be dominated by the rare {\it
rock} events, an experiment which aims at exploiting this effect
should maximize the acceptance for almost horizontal events. We
conclude by noticing that our results do not take into account any
detailed experimental setup and up to this point the $\nu_\mu$
contribution is not yet considered. Nevertheless, since both the
incoming neutrino flux and cross section on nucleons are expected to
be flavor independent the possibility of determining both these
quantities at a NT seems an interesting perspective. A more detailed
analysis of this issue will be addressed elsewhere.

%%%%%%%%%%%%%%%%%%%%%%%%%%%%%%%%%%%%%%%%%%%%%%%%%%%%%%%%%%%%%%%%%
\section{The $\nu_\mu$ contribution: disentangling $\mu$'s from $\tau$'s}
\label{nu_mu}

In the previous Section we have discussed the rate of $\nu_{\tau}$
events implicitly assuming that a $\tau$ lepton can be
distinguished from a $\mu$ in a NT. However, given the
experimental characteristics of the detector, this could be a
difficult task and $\nu_{\mu}$ events should be also included in
any realistic simulation. We address this issue in the present
Section.
\begin{figure}[!tb]
\begin{center}
\begin{tabular}{cc}
\epsfig{figure=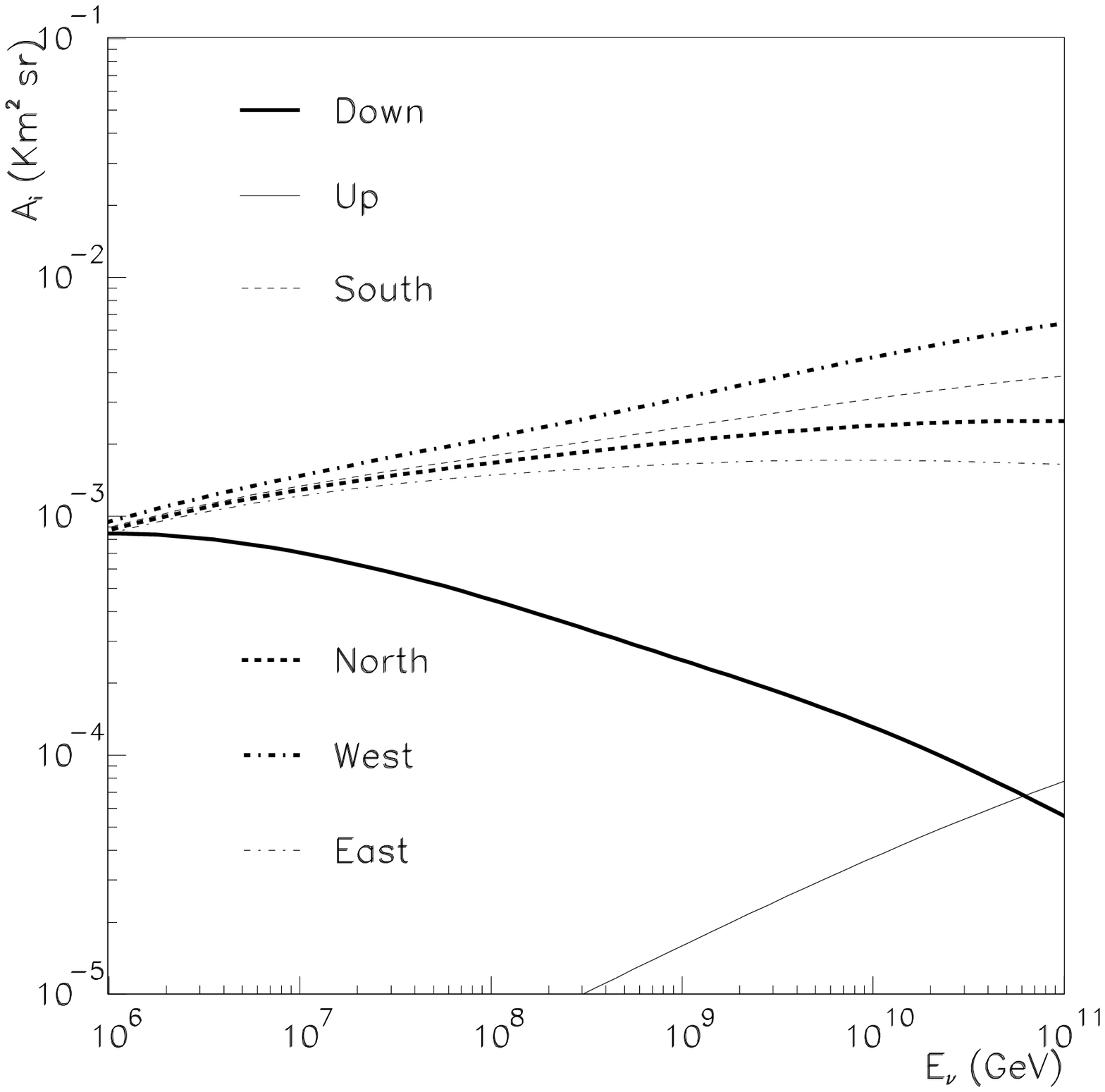,height=7cm} &
\epsfig{figure=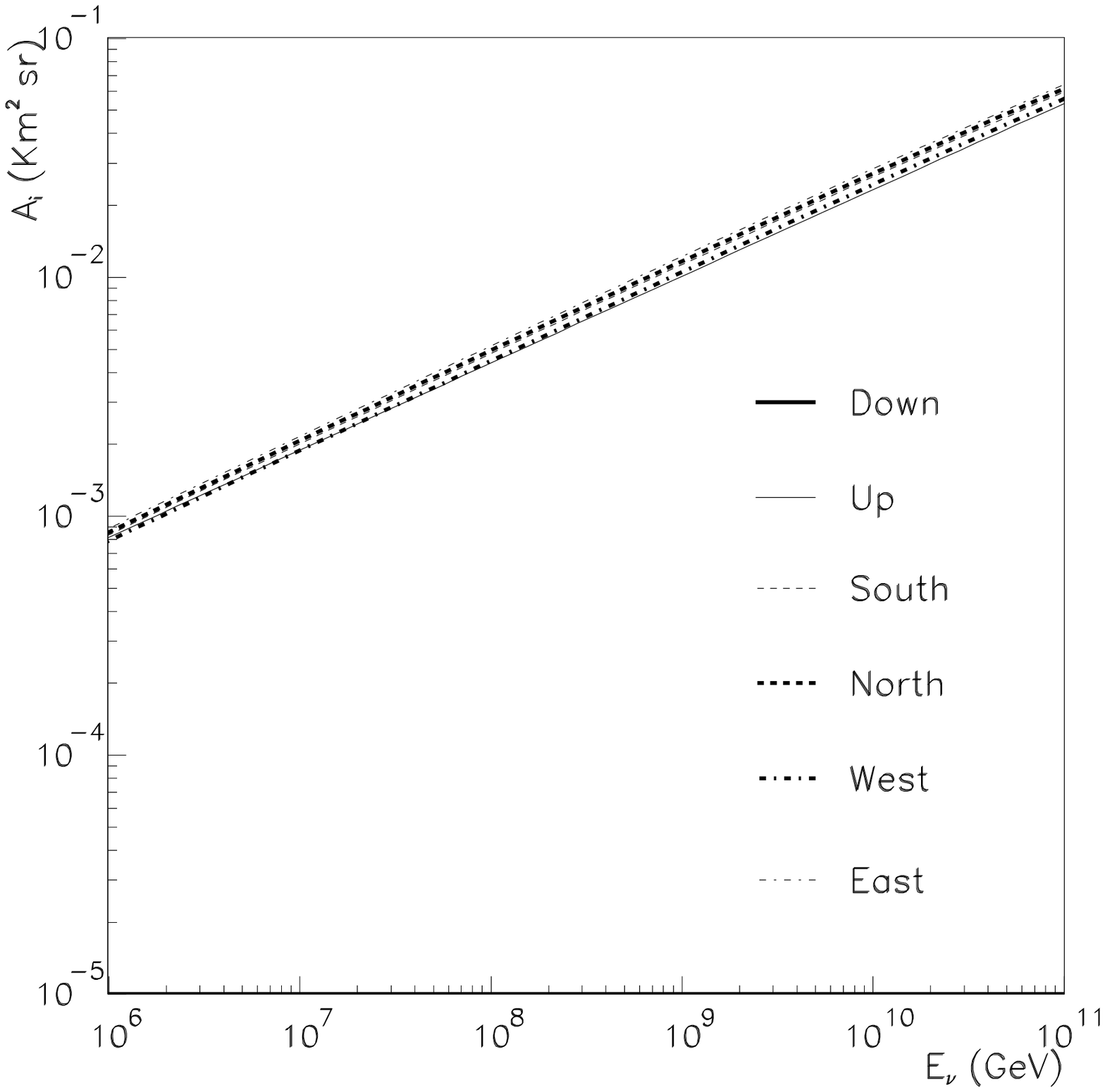,height=7cm}
\end{tabular}
\end{center}
\caption{The effective apertures $A_a^{\mu(r,w)}(E_\nu)$ versus the
muon neutrino energy for (left) {\it rock} events and (right) {\it
water} events at the  \texttt{NEMO} site. } \label{apertsurfmu}
\end{figure}

Using the definition of the aperture in full analogy with Eq.\
(\ref{kernel2}) and applying the same considerations of Section 2 to
$\nu_\mu/\mu$, we have computed the $\mu$ apertures for {\it water}
and {\it rock} events for the various surfaces of the NT, adopting
the value $\beta_\mu=0.58 \times 10^{-5}$ cm$^2$ g$^{-1}$ in the
expression of the muon differential energy loss analogous to Eq.\
(\ref{dedz}) (as for the $\tau$, the term weighted by $\gamma_\mu$
is negligible for the energy range of interest). The results are
shown in Fig.\ \ref{apertsurfmu}. The main features as well as the
role of matter effects are essentially unchanged for muons, the only
difference coming from the muon contribution at lower energies
because of the longer muon lifetime compared to that of taus.

It is worth to discuss briefly the adopted parametrization for the
muon energy losses. As discussed above, tau energy losses are
affected by the large theoretical uncertainty on cross-section for
photonuclear interaction. On the other hand, photonuclear
interactions are less relevant for muon propagation, thus the
theoretical uncertainty on the energy loss is correspondingly
smaller. We then checked the validity of the approximation given
in Section~\ref{nu_mu} versus the detailed calculation given in
\cite{Bottai:2000en}. We found that the accuracy is at the level
of 15\% over the whole energy range. The impact of this
uncertainty on the expected event rate is as follows: a 15\%
increase of $\beta_\mu$ gives a  few \% decrease of the number of
water events and a $\sim$ 10 \% decrease of the number of rock
events. This uncertainty then does not affect the estimate of the
number of $\nu_{\mu}$ events while a more careful treatment is
required for a reliable forecast of the neutrino cross-section
sensitivities at a Neutrino Telescope.
\begin{figure}[!tb]
\begin{center}
\begin{tabular}{cc}
\epsfig{figure=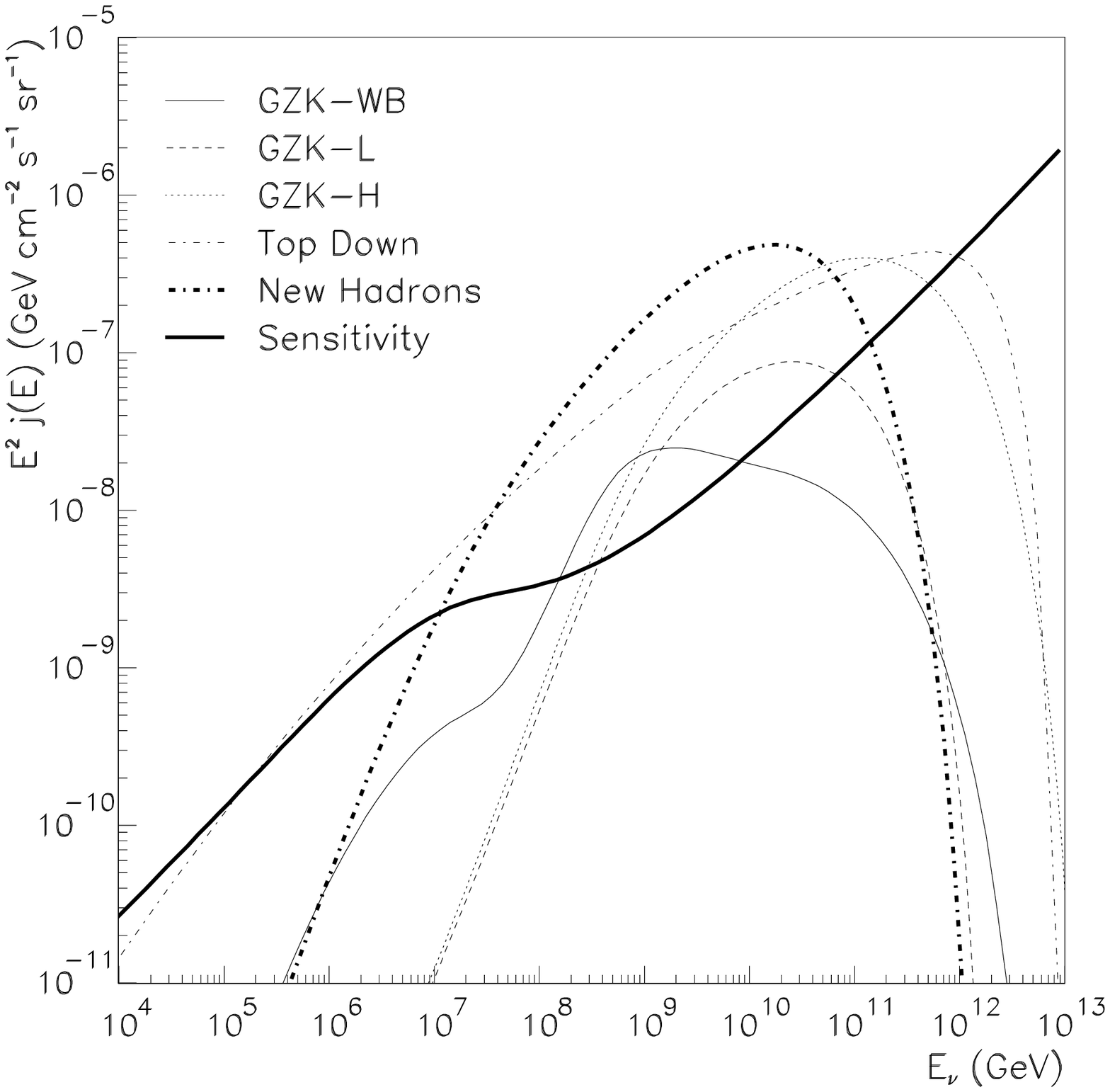,height=7cm} &
\epsfig{figure=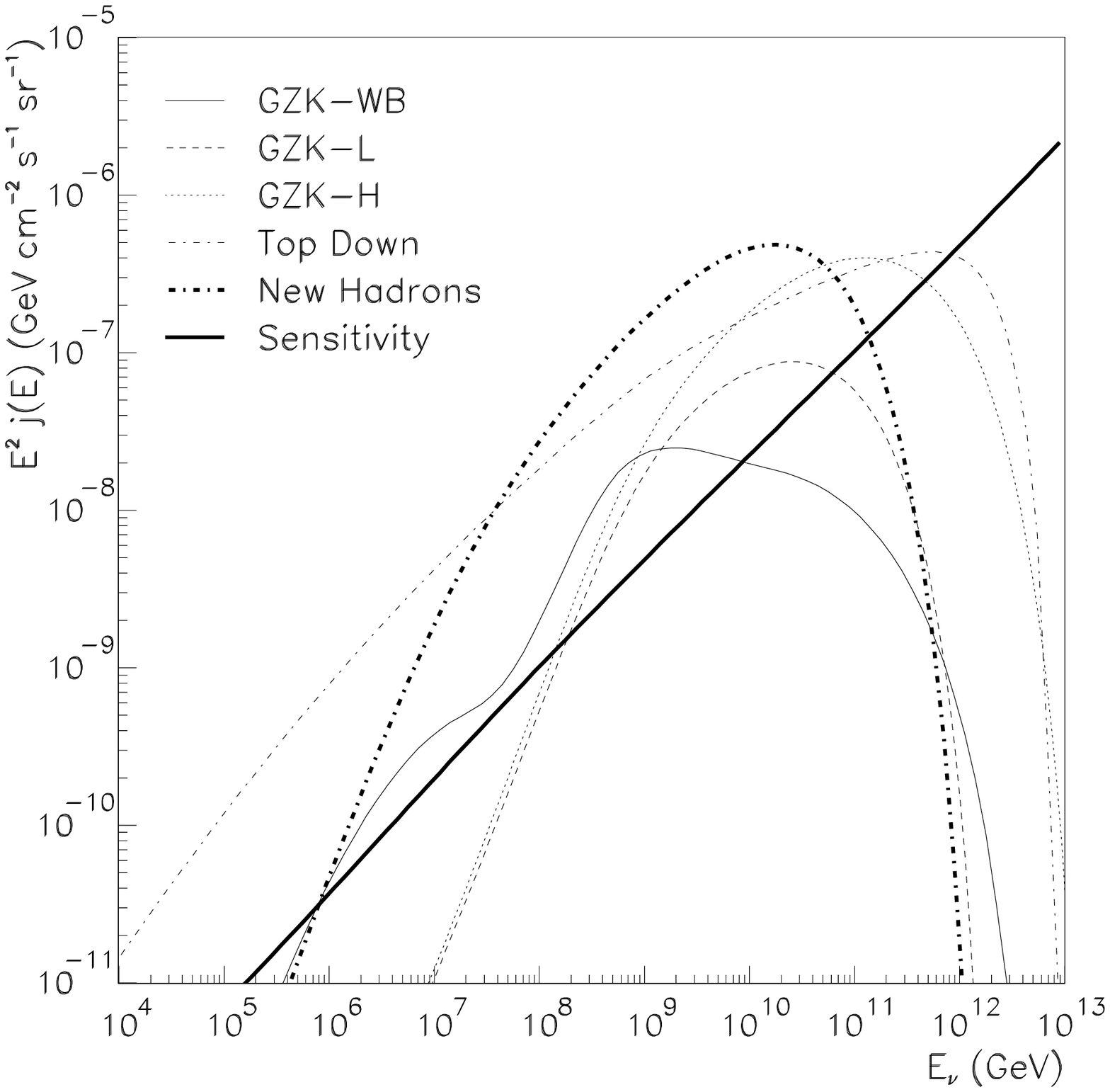,height=7cm}
\end{tabular}
\end{center}
\caption{The total \texttt{NEMO} sensitivities for (left) $\tau$
and (right) $\mu$ events versus the neutrino energy, compared with
the UHE neutrino fluxes considered in the paper.}
\label{sensitivities}
\end{figure}

In Fig. \ref{sensitivities} we summarize the $\tau$ and $\mu$
results by showing the total sensitivity $S^{\mu,\tau}$ for
\verb"NEMO" defined as $S^{\mu,\tau}\, E_\nu \, A^{\mu,\tau}= 1$
event year$^{-1}$ ($E_\nu$ decade)$^{-1}$, with $A^{\mu, \tau}$ the
total $\mu$ and $\tau$ effective aperture, respectively; we also
show the various neutrino fluxes considered through the paper. We
see that in agreement with the results of Table \ref{table::events},
at least one event per year is expected even in the case of a GZK-WB
flux, while larger rates are expected for higher fluxes (see also
\cite{Barbot:2002kh}). Notice that in the energy bin $10^8-10^{10}$
GeV both $\mu$ and $\tau$ contributions are comparable while muons
are expected to dominate in the lower energy range, depending on the
particular flux we consider.

\begin{figure}[t]
\begin{center}
\epsfig{figure=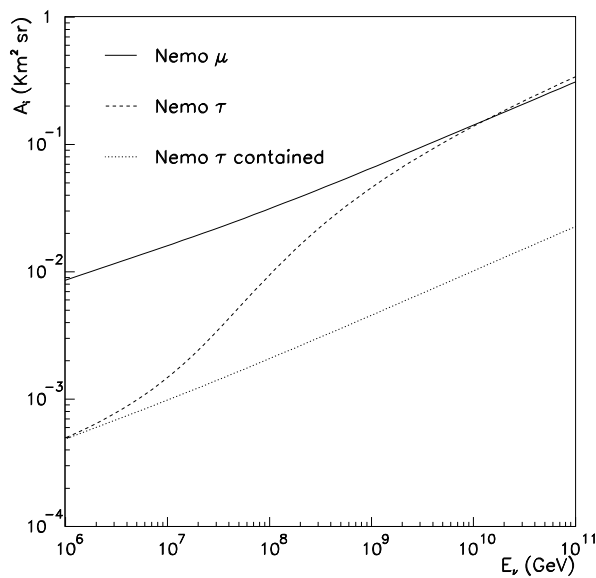,height=8cm} \caption{Total $\tau$,
$\mu$ and contained events apertures for the \texttt{NEMO} site.
Note that the contained events are the same for all neutrino
flavors.} \label{nemocontained}
\end{center}
\end{figure}

Concerning the possibility to distinguish between UHE taus and
less energetic muons a comment is in turn. As we mentioned in the
Introduction, the main difficulty is that Cherenkov detectors like
a NT do not measure the particle energy but rather the energy loss
inside the detector volume and thus a high-energy $\tau$ tracks
can be misidentified with a muon track of lower energy. This is
because the ratio of $\tau$ to $\mu$ energy loss rate is given by
$\beta_{\tau}/\beta_{\mu}\simeq 1/8$. In principle, given that tau
energy losses are dominated by photonuclear processes versus
radiative interactions for muons, it would be possible to
distinguish a muon-track from a tau-track from the different
hadronic content along the track. However, Montecarlo simulations
indicate that NT's are poorly sensitive to this signature
\cite{Bugaev:2003sw}.

As far as the contained events are concerned (i.e. where the
charged lepton production happens inside the instrumented volume),
the telescope has, instead, realistic chances of flavor-tagging.
As long as the energy-loss or decay range of the particle to be
detected is small compared with the detector size, the event rate
depends basically on the fiducial volume. This is always the case
for neutral current events, which if detectable produce a
localized hadronic shower from the struck nucleon, and for charged
current $\nu_e$-events, since the electron rapidly loses its
energy. On the contrary, for energies above the TeV scale,
$\nu_\mu$'s charged current events produce muons which are
detectable as tracks several km away from the production point.
The case of $\nu_\tau$ charged current events is yet different:
for energies $\alt 10^7$ GeV, the boosted decay range of the tau
particle is negligible with respect to the detector size and
depth, and the event rate is determined by the instrumented
volume, like for $\nu_e$-events. On the other hand, for the
typical spacings between strings/photomultipliers considered by
current designs, above the PeV scale the boosted $\tau$ decay
length is larger than the spacing and one starts resolving the
{\it tracks}, while below this energy $\tau$ produce showers which
differ from $\nu_e$-events only by the hadronic content, which is
however challenging to tag.

We show in Fig. \ref{nemocontained} the total apertures for $\tau$
and $\mu$ events at the \verb"NEMO" site together with the aperture
for $\tau$ contained events (the same would apply for $\mu$
contained events). Indeed, contained events represent a sub-leading
although non-negligible fraction of the total number of events,
always of the order of 10\% for $\mu $ and even greater for low
energy $\tau$ ($\alt 10^7$ GeV) due to the very short ($\alt 1$ km)
decay length at these energies. Moreover, the contained aperture
depends only on the neutrino interaction probability so that it can
be considered also a reliable estimate of the $e$-induced showering
events, neglecting the effect of the Glashow $\bar{\nu}_ee^-$
resonance at 6.3 PeV. It may also be possible to identify {\it
lollipop} events in which a $\tau$ with energy larger than PeV
produces a long minimum-ionizing track that enters the detector and
eventually ends in a huge burst as the $\tau$ lepton decays into a
final state with hadrons or an electron. In this case, the final
burst would be a direct measurement of the $\tau$ energy while the
energy loss along the track would be smaller than for a muon of the
same energy. Probably, the cleanest signature of a $\tau$ particle
would be the detections of a double-bang event \cite{Learned:1994wg}
in which a $\nu_{\tau}$ interacts inside the detector and the
produced $\tau$ decays in shower again in the detector, but the
probability of such an event is extremely small. In any case, all
these possibilities suffer from lower statistics as they all require
that the interactions (showering or production) occur inside the
detector, with a reduction of the effective volume down to 1 km$^3$
compared with the several km$^3$ effective volume for $\mu$ and
$\tau$ events which go across the NT fiducial volume.

\begin{figure}[!tb]
\begin{center}
\begin{tabular}{cc}
\epsfig{figure=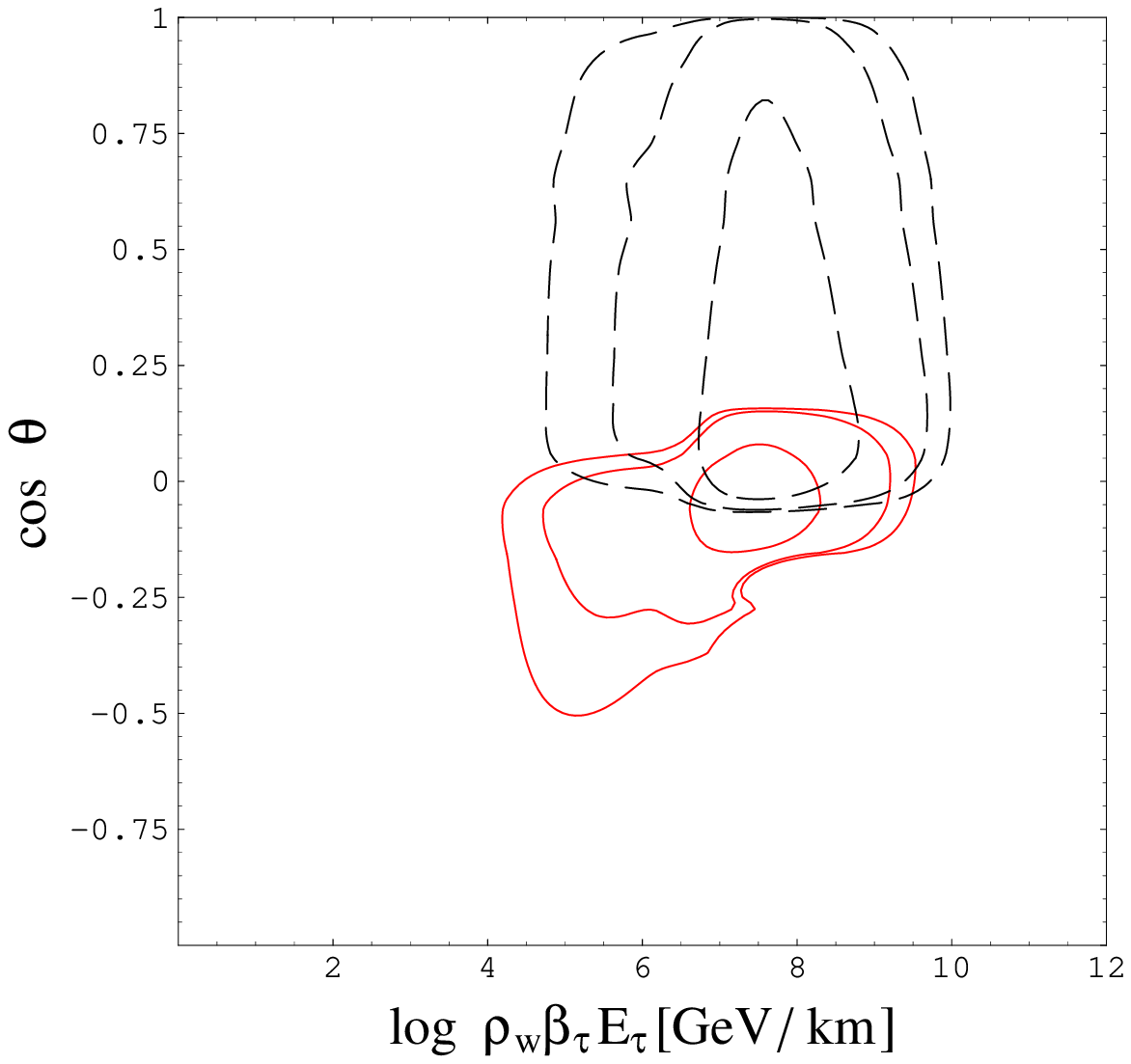,height=7cm} &
\epsfig{figure=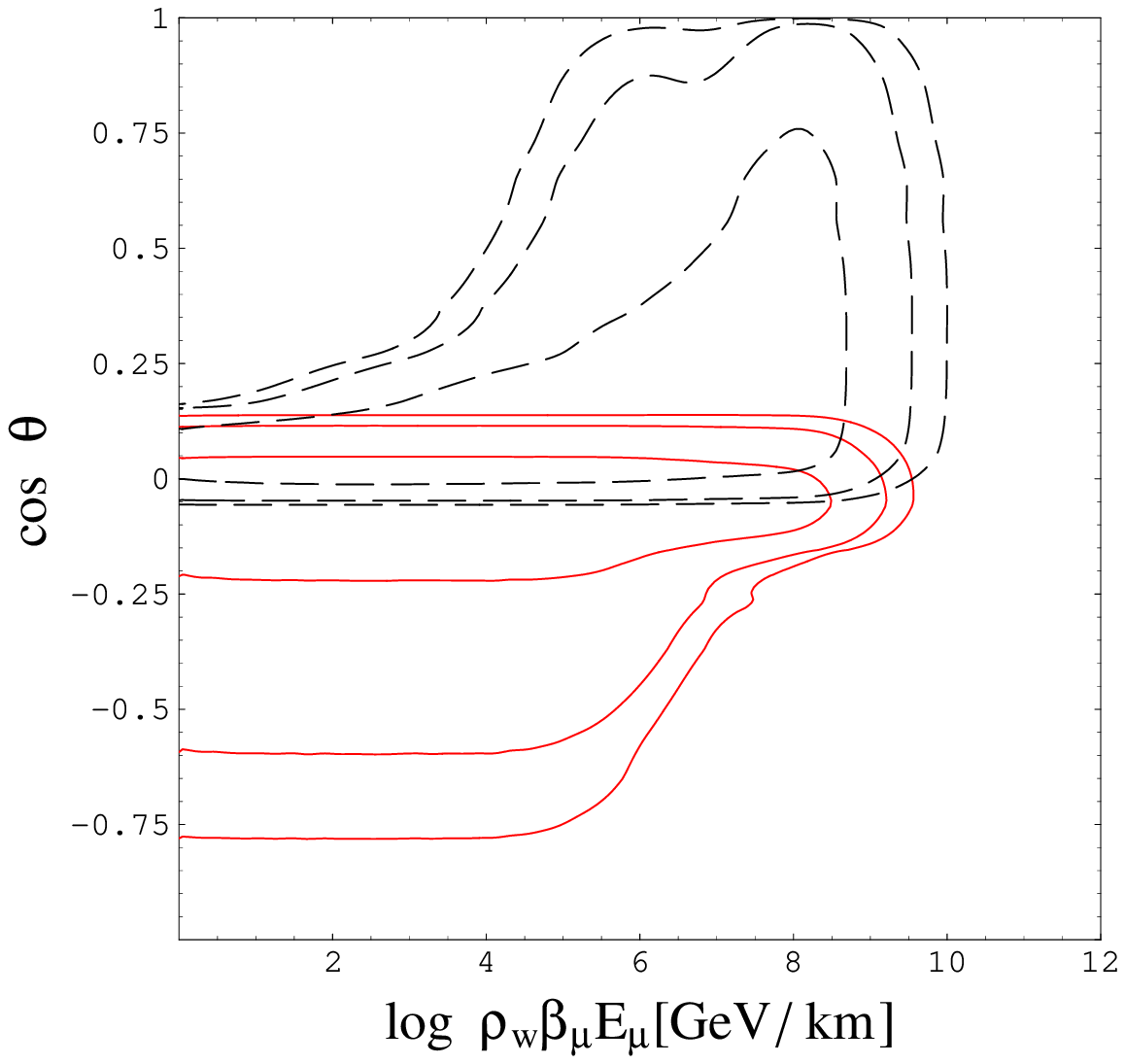,height=7cm}
\end{tabular}
\end{center}
\caption{Contour plots of the number of {\it rock} events (red
full lines) and {\it water} events (black dashed lines) at the
\texttt{NEMO} site in the zenith angle-$\d E/\d x$ plane for
$\tau$ (left panel) and  $\mu$ (right panel) assuming a GZK-WB
neutrino flux. The contours enclose 65, 95 and 99 $\%$ of the
total number of events.} \label{ratecontours}
\end{figure}

In view of these considerations, we conclude that, at least for
the bulk of the events, the most viable strategy is to combine
both muon and tau contributions and construct spectra depending
upon quantities which are directly observable. The simplest choice
is to consider the energy loss rate inside the detector which
amounts to measure the track length and the total deposited energy
and the arrival direction. In Fig. \ref{ratecontours} we show the
contours of expected $\mu$ and $\tau$ event rates in terms of the
zenith angle of arrival directions and $\d E/\d x\simeq - \beta
E\varrho_{w} $. For relatively low energy losses $\beta
E\varrho_{w}\alt 10^5\,$GeV/km the whole dominant contribution
comes from muons, which therefore can be easily disentangled,
whereas in the high energy loss tails the event distributions are
almost the same for both neutrino flavors and one is forced to use
the total $\nu_\mu$ + $\nu_\tau$ events as the input of any
analysis of the data.
%

%%%%%%%%%%%%%%%%%%%%%%%%%%%%%%%%%
\section{Conclusions}\label{concl}

Ultra-high energy neutrinos represent one of the main targets for
several experiments which adopt a variety of detection techniques.
Among these, the optical Cherenkov neutrino telescopes deployed
under water or ice look for the tracks of charged leptons produced
by the high energy neutrinos that reach the Earth. In this paper, we
have presented a new study of the performance of a km$^3$ neutrino
telescope to be located in any of the three sites in the
Mediterranean sea proposed by the \texttt{ANTARES}, \texttt{NEMO},
and \texttt{NESTOR} collaborations. Our main goal is to compare the
performances of the different sites, keeping apart detector-specific
features, like partial detection efficiency or the architecture
adopted for the towers of strings. We concentrated instead on the
details of the under-water surface profile of each of the three
sites, using the data from a Digital Elevation Map. By generating a
realistic and statistically significant sample of $\nu_\tau$/$\tau$
and $\nu_\mu$/$\mu$ tracks crossing the fiducial volume of the
km$^3$ neutrino telescope, we have calculated its effective aperture
to UHE $\nu_\tau$ and $\nu_\mu$ neutrinos and the expected number of
events for different UHE neutrino fluxes, for both cases where the
neutrino/charged lepton track is crossing the rock (denoted as {\it
rock events}) or the water only (denoted as {\it water events}). Our
results can be summarized as follows:
\begin{itemize}
\item The impact of the site geography
(or matter effects) on observables such as the ``rock fraction" of
the total event rate  or asymmetries in the event direction can be
important, particularly at high energies.

\item Even for a fixed instrumented volume, these matter effects can be
enhanced by a suitable choice of the geometry of the telescope,
maximizing the lateral surface of the fiducial volume.

\item The continental crust provides an absolute
orientation, hence the matter effects may provide a mean for
calibrating the pointing capabilities of the detector, even when
no point-source identification is possible, like for diffuse
cosmic fluxes at UHE.

\item  We analyzed briefly the dependence of the rock and water events
from the neutrino fluxes and the neutrino-nucleon cross section. We
found that the ratio of rock to water events may provide an
additional way to disentangle the two unknowns, in addition e.g. to
the well-known zenith angle dependence of the rates due to the
screening effect of the average spherical Earth. Although a detailed
analysis would be needed, we stress that this may be important for
constraining neutrino-nucleon cross section at UHE energies,
otherwise unaccessible at the Lab.

\item While below the PeV scale the aperture for muon tracks is one
order of magnitude larger than the aperture for contained events,
above $\sim 10^8$ GeV the numbers of muon and tau track events are
comparable. We have briefly addressed the problem of whether it is
possible to distinguish $\mu$'s and $\tau$'s in the detector.
Apart for the sub-dominant fraction of contained events with
specific signatures, we stressed that a realistic prescription at
UHE is to sum the bulk of $\mu$ and $\tau$ events, the natural
variables for describing the events being the arrival direction
and the energy loss rate in the fiducial volume.
\end{itemize}

The main conclusion one can draw from our analysis is that the
optimization for a telescope aiming at the $E>$PeV region is
significantly different from one whose target is the $E\sim$TeV
range: in the first case, the search is basically background free
and even a relatively poor angular and energy resolution may be
acceptable. The crucial goal is to maximize the event rates, and
the discrimination among models may be based on ``counts" and a
very rough directional and energy binning of the events. In this
respect, one should maximize the instrumented volume---compatibly
with the experimental requirements for a meaningful reconstruction
of the event---and also carefully design the geometry to maximize
the lateral surface, in order to exploit the matter effects
provided by the underwater profile. On the other hand, at TeV
energies angular and energy resolutions are crucial to improve the
very signal to noise ratio, and may help identifying point-like
sources. At the same time, in the TeV range the matter effects we
have stressed on in this paper are less relevant, and should not
influence significantly the choice of the site.

%%%%%%%%%%%%%%%%%%%%%%%%%%%%%%%%%%%%%%%%%%%%%%%%%%%%%%%%%%%%%%%%%%%%%%
%% Acknowledgments %%%%%%%%%%%%%%%%%%%%%%%%%%%%%%%%%%%%%%%%%%%%%%%%%%%
%%%%%%%%%%%%%%%%%%%%%%%%%%%%%%%%%%%%%%%%%%%%%%%%%%%%%%%%%%%%%%%%%%%%%%
\section*{Acknowledgments}
We are pleased to thank G. Barbarino, D. Hooper, G. Longo, and E.
Migneco for useful comments and discussions. This work was supported
by a Spanish-Italian AI, the Spanish grants FPA2005-01269 and
GV/05/017 of Generalitat Valenciana, a MEC-INFN agreement as well as
the PRIN04 "Fisica Astroparticellare" of Italian MIUR. SP was
supported by a Ram\'{o}n y Cajal contract of MEC. PS acknowledges
the support by the Deut\-sche For\-schungs\-ge\-mein\-schaft under
grant SFB 375 and by the European Network of Theoretical
Astroparticle Physics ILIAS/N6 under contract number
RII3-CT-2004-506222.

%%%%%%%%%%%%%%%%%%%%%%%%%%%%%%%%%%%%%%%%%%%%%%%%%%%%%%%%%%%%%%%%%%%%%%%
\section*{References}
%%%%%%%%%%%%%%%%%%%%%%%%%%%%%%%%%%%%%%%%%%%%%%%%%%%%%%%%%%%%%%%%%%%%%%%

\end{document}